\documentclass[english,aps,twocolumn,nofootinbib,showpacs,longbibliography,secnumarabic,superscriptaddress,floatfix]{revtex4}

\usepackage[english]{babel}
\usepackage{amssymb,amsmath,soul}

\usepackage[pdftex]{graphicx}
\usepackage[caption=false,subrefformat=parens,labelformat=parens]{subfig}

\usepackage{color}
\usepackage[usenames,dvipsnames]{xcolor}
\usepackage{hyperref} 
\hypersetup{%
	linktoc=page,%
	colorlinks=true,%
	linkcolor=RoyalBlue,%
	citecolor=OliveGreen,%
	urlcolor=OliveGreen,%
	linkbordercolor=red,%
	pdfborderstyle={/S/U/W 1},%
}
\setul{1.4pt}{.4pt}

\usepackage{grffile}
\usepackage{balance}
\usepackage{float}

\newcommand{\uhref}[2]{{\href{#1}{\ul{#2}}}}

\renewcommand{\vec}{\boldsymbol}
\newcommand{\C}{\bar{C}}
\newcommand{\sss}{\scriptscriptstyle}

\newcommand{\IV}{$I\mbox{-}V$}

\newcommand{\up}{u_{\sss \! \perp}}
\newcommand{\eone}{\varepsilon_{1}}

\begin{document}

\title{Strong-pinning regimes by spherical inclusions in anisotropic type-II superconductors}

\author{Roland Willa}
\affiliation{
	Materials Science Division,
	Argonne National Laboratory,
	9700 South Cass Avenue, Lemont, Illinois 60439, USA
}

\author{Alexei E. Koshelev}
\affiliation{
	Materials Science Division,
	Argonne National Laboratory,
	9700 South Cass Avenue, Lemont, Illinois 60439, USA
}

\author{Ivan A. Sadovskyy}
\affiliation{
	Materials Science Division,
	Argonne National Laboratory,
	9700 South Cass Avenue, Lemont, Illinois 60439, USA
}
\affiliation{
	Computation Institute, 
	University of Chicago, 
	5735 S. Ellis Av., Chicago, Illinois 60637, USA
}

\author{Andreas Glatz}
\affiliation{
	Materials Science Division,
	Argonne National Laboratory,
	9700 South Cass Avenue, Lemont, Illinois 60439, USA
}
\affiliation{
	Department of Physics,
	Northern Illinois University,
	DeKalb, Illinois 60115, USA
}

\begin{abstract}
The current-carrying capacity of type-II superconductors is decisively determined by how well material defect structures can immobilize vortex lines. In order to gain deeper insights into the fundamental pinning mechanisms, we have explored the case of vortex trapping by randomly distributed spherical inclusions using large-scale simulations of the time-dependent Ginzburg-Landau equations. We find that for a small density of particles having diameters of two coherence lengths, the vortex lattice preserves its structure and the critical current $j_c$ decays with the magnetic field following a power-law $B^{-\alpha}$ with $\alpha \approx 0.66$, which is consistent with predictions of strong-pinning theory. For a higher density of particles and/or larger inclusions, the lattice becomes progressively more disordered and the exponent smoothly decreases down to $\alpha \approx 0.3$. At high magnetic fields, all inclusions capture a vortex and the critical current decays faster than $B^{-1}$ as would be expected by theory. In the case of larger inclusions with a diameter of four coherence length, the magnetic-field dependence of the critical current is strongly affected by the ability of inclusions to capture multiple vortex lines. We found that at small densities, the fraction of inclusions trapping two vortex lines rapidly grows within narrow field range leading to a peak in $j_c(B)$-dependence within this range. With increasing inclusion density, this peak transforms into a plateau, which then smooths out. Using the insights gained from simulations, we determine the limits of applicability of strong-pinning theory and provide different routes to describe vortex pinning beyond those bounds.
\end{abstract}

\pacs{
	74.25.Wx,	% Vortex pinning (includes mechanisms and flux creep)
	74.25.Sv,	% Critical currents
	74.20.De,	% Phenomenological theories (two-fluid, Ginzburg-Landau, etc.)
	74.25.Uv	% Vortex phases (includes vortex lattices, vortex liquids, and vortex glasses)
}

\maketitle

\onecolumngrid
\tableofcontents
\clearpage
\twocolumngrid

\section{Introduction} \label{sec:introduction}

Magnetic flux-lines, or vortices, in type-II superconductors represent a unique exemplary system perfectly suited for studying periodic structures driven through a quenched random potential. The development of quantitative descriptions for vortex systems poses a long-standing challenge, the importance of which cannot be overemphasized: emerging high-current applications of superconductors strongly rely on efficient immobilization of flux lines by artificially-created defect structures. Incorporating self-assembled inclusions into high-temperature superconductors has been established as a very efficient route to enhance their critical currents. Depending on the fabrication process, these inclusions may be prepared in the form of almost spherical particles~\cite{MacManusAPL04, HauganNat04, GutierrezNatMat07, YamasakiSUST08, PolatPhysRevB11, MiuraPhysRevB11, MiuraSUST13, MeleSuST15, Haberkorn2017, JhaJAP17}, nanorods~\cite{GoyalSUST05, KangSci06}, or combinations thereof~\cite{MaiorovNaMat09, JhaSuST15}. This technology is implemented in today's second-generation superconducting wires based on rare-earth barium copper oxide (REBa$_{2}$Cu$_{3}$O$_{7}$ or REBCO) coated conductors~\cite{MalozemoffAnnRevMatRes12, ObradorsSST2014}, where the rare earth (RE) is mostly yttrium (Y) or gadolinium (Gd). More recently, similar approaches have been used to enhance pinning in another family of high-performance superconductors, namely in iron pnictides~\cite{Tarantini2012, MiuraNComm2013}.  

Despite the development of analytic models to capture the vortex dynamics through random disorder potentials, the complicated pinning landscape found in state-of-the-art superconductors remains out of reach for analytical descriptions. The rational optimization of pinscapes may then be facilitated by investigating vortex pinning with large-scale numerical simulations, laying a foundation for the critical-current by design paradigm~\cite{SadovskyyAdMat2016}. In reciprocity, the gained insights may allow for a better understanding of the vortex pinning/depinning mechanisms.

Ultimately, the route towards largest possible critical currents lies in the constructive combination of different pinning centers. A natural first step on this journey consists of finding the optimal pinning configuration for a relatively simple model system with only one type of defects. In this work we limit ourselves to monodisperse spherical inclusions. While having in mind self-assembled nanoparticles in coated REBCO conductors, similar pinning centers---in the form of impurity clusters introduced by proton or ion irradiation---are known to further enhance the critical current in these materials~\cite{Matsui2012, JiaAPL13, Haberkorn2015b, LerouxAPL2015}, as well as in iron-based superconductors~\cite{TaenPRB2012, HaberkornPRB2012, KihlstromAPL2013, TaenSST2015}.

Even for such simple model systems, a \emph{quantitative} description of the vortex dynamics, e.g., predicting the dependences of the critical current on the magnetic field as well as on the density and strength of pinning centers, poses a difficult problem. Indeed, vortex pinning is a complex collective phenomenon controlled by (i)~the interaction of vortices with pinning sites, (ii)~the elastic properties of the flexible vortex lines, and (iii)~their mutual interactions. In the case of weak pinning by a large density of atomic impurities~\cite{LarkinO1979}, the analytical treatment of this problem is limited to qualitative estimates, providing scaling laws for the critical current. The situation improves when pinning is produced by a dilute distribution of strong defects interacting with an ordered vortex lattice~\cite{OvchinnikovI1991, BlatterGK2004}.  In this case, the calculation of macroscopic quantities such as the critical current, or the Campbell length can be done at a quantitative level~\cite{BlatterGK2004, Thomann2012, Willa2015a, Willa2016}. Both pinning cases have been discussed in detail in several reviews~\cite{BlatterFGLV1994, Brandt1995, BlatterG2008, GurevichAnnRevCMP14, KwokRoPP2016}. Despite the advantage of the strong-pinning formalism over the weak collective theory in the ability to classify pinning regimes, it should be noted that both approaches inevitably rely on simplifying assumptions and thereby miss important details.

The idea that numerical routines may give a more realistic insight into the dynamics of flux lines is not new~\cite{BrandtJLTP83-1, BrandtJLTP83-2}: over the last decades, several approaches have been used to model vortex states in superconducting materials. In the minimal approach, the problem is reduced to vortex degrees of freedom only. Hence, vortices are treated as particles (in thin films) or elastic strings (in bulk) and their dynamics is described by an overdamped equation of motion, which takes into account interaction with pinning centers and the thermal Langevin forces. This Langevin-dynamics approach provides a qualitative description of the vortex state in small fields, when the distance between vortices is much larger than the coherence length, and for small density of pinning centers. In particular, for the three-dimensional case, such simulations have been used to explore the vortex dynamics in Refs.~\cite{ErtasK1996, BustingorryCD2007, LuoHu2007, KoshelevK2011, DobramyslEPJ2013, AssiPRE16}. Due to a minimum number of degrees of freedom explored, this simple and physically transparent approach allows studying large systems with good statistics. This description, however, has several limitations: vortex-vortex and vortex-pin interactions can only be treated approximately, and the possibilities of vortex cutting and reconnection are completely neglected. Furthermore, this model fails to properly treat the most relevant case when pinning centers occupy a noticeable fraction of the superconducting volume. It is therefore desirable to probe the strong-pinning regime within a more sophisticated model.

All aforementioned limitations are overcome in the time-dependent Ginzburg-Landau (TDGL) model~\cite{Schmid1966} describing the superconducting order parameter in a driven state. At finite  magnetic fields,  vortex lines appear spontaneously as singularities in the phase of the complex order parameter. Despite its physical transparency, the TDGL model is also subject to several limitations regarding a realistic description of the vortex dynamics. Notwithstanding this note of caution, the TDGL model is well suited for studying static pinning problems, where an accurate description of dynamics is not essential. In the past, the TDGL model has proven itself to be very useful for exploring numerous properties of the vortex state~\cite{Doria1990, Machida1993, Crabtree1996, Aranson1996, Crabtree2000, WinieckiA2002, Vodolazov2013, Berdiyorov2014}. Recent developments of a high-performance, parallel TDGL solver~\cite{SadovskyyJComp2015}, enabled the meaningful exploration of the parameter space for sufficiently large three-dimensional superconductors. This solver, implemented for GPUs, has been used to tackle various problems, including the study of pinning in realistic sample geometries, which reconstructed from a 3D STEM tomogram of {Dy-doped} YBCO~\cite{SadovskyyPRAppl2016}, vortex dynamics in ordered and hyperuniform patterned thin films~\cite{SadovskyyPRB2017}, the process of vortex cutting and reconnection~\cite{GlatzVKC2016,vlasko+prb15}, the effect of geometrical pinning in nanobridges~\cite{Papari2016}, and the optimization of pinning configurations~\cite{KoshelevPRB16, SadovskyyPRE2017}, see also Ref.~\cite{KwokRoPP2016}.

In this paper, we explore the regimes of strong vortex pinning within the TDGL framework. For this purpose, we investigate the pinning capability of a low density of strong defects (in our case spherical, normal inclusions). Contrasting the numerical results with existing theoretical predictions will provide limits of applicability of the latter. At the same time, the computational efforts will provide useful feedback to improve the analytical description beyond today's limits which, in turn, will facilitate a better interpretation of experimental data.

The paper is organized as follows. In Sec.~\ref{sec:estimates} we review established analytical results for the critical current at small pin densities and their limits of applicability. Supported by our numerical results, we discuss various approaches to go beyond those limits. A brief description of the TDGL model used for our numerical calculations is given in Sec.~\ref{sec:model} (for details on the technical realization of the numerical solver, we refer the reader to Ref.~\cite{SadovskyyJComp2015}). In order to characterize the properties of the elemental contributor to vortex pinning, we investigate isolated inclusions with TDGL simulations in section \ref{sec:single_inclusion}. In Sec.~\ref{sec:field_dependence}, we study  the dependence of the critical current on the magnetic field strength and inclusion density in detail for two different particle sizes (two and four coherence lengths in diameter). Different parameters, extracted from the simulations, help us better understand and quantify the mechanisms of vortex pinning. For this analysis, field-induced vortex lines are extracted from the complex order-parameter function by means of a routine described in Ref.~\cite{PhillipsPRE2015}. The numerical results are compared with theoretical expectations.

\section{Strong-pinning theory for different magnetic-field regimes} \label{sec:estimates}

\begin{figure}[tb]
	\centering
	\includegraphics[width=0.47\textwidth]{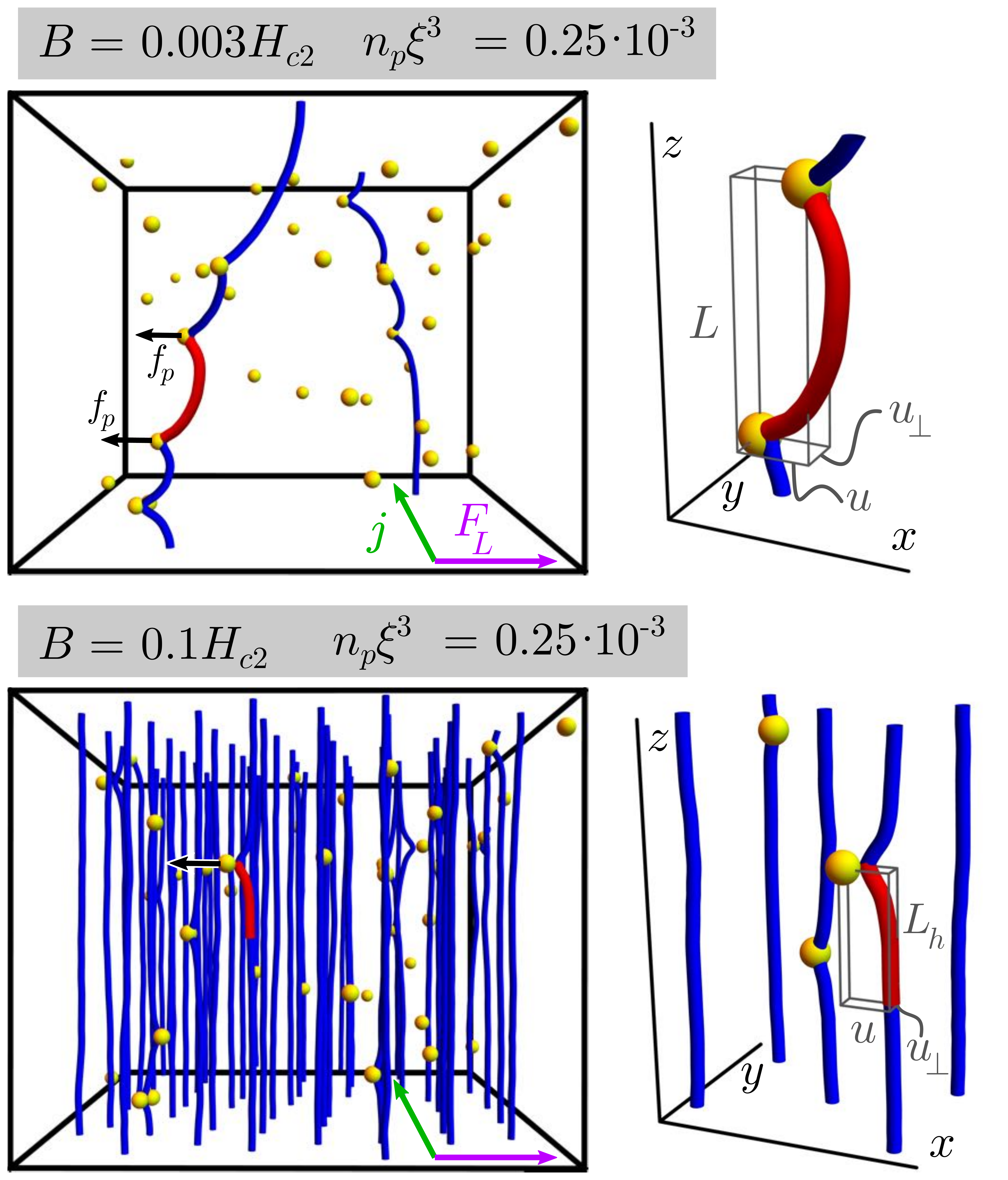}
	\caption{
		Illustration of pinned vortex configurations for the single-line (1D) 
		and lattice (3D) regimes obtained using TDGL simulation. 
		The extent of length scales $u$, $\up$, $L$, and $L_{h}$ 
		defines the relevant pinning volumes, shown as gray boxes.
	}
	\label{fig:vortices_1d_3d}
\end{figure}

The theory of strong vortex pinning describes the interaction of vortices with a low density $n_{p}$ of strong defects.\footnote{This \textit{strong-pinning} regime has to be contrasted to the \textit{weak collective limit} where only fluctuations of defect density  provide a finite pinning force on the vortex system} The defect strength guarantees that each inclusion is capable of pinning a vortex even if isolated from all the others~\cite{Labusch1969}. In the low defect density limit, the problem reduces to pinning of vortex segments trapped between two defects. These trapped segments are characterized by their typical length $L$ along the field direction $z$, displacement $u$ in the direction $x$ of vortex motion, and displacement $\up$ transverse to that motion (along~$y$), see Fig.~\ref{fig:vortices_1d_3d}. Since each of these vortex segments is unpinned inside a volume $L u \up$, the three lengths are related through the geometric constraint
\begin{equation}
	n_{p} L u \up \approx 1.
	\label{eq:geom-relation}
\end{equation}
Let $f_{p}$ denote the maximal pinning (or pin-breaking) force an isolated defect can exert on the vortex line. The critical current $j_{c}$ necessary to detach the vortex from the pinning site is then determined by the length of the trapped segment and the pin-breaking force $f_{p}$ via
\begin{equation}
	\frac{\Phi_{0}}{c}j_{c}\approx \frac{f_{p}}{L}.
	\label{eq:CritCurrTrapSegm}
\end{equation}
While the pin-breaking force $f_{p}$ is mostly a property of the defect (at least, for small magnetic fields), the typical segment length $L$ results from the complex interplay between the vortex-pin interaction, the line tension, and the interactions between different vortex lines~\cite{OvchinnikovI1991, BlatterGK2004, KoshelevK2011}. In the following subsections we will review specific cases of the strong-pinning theory.

\subsection{1D strong-pinning theory} \label{sec:1D}

At very small magnetic fields, the interaction between vortices is irrelevant and flux lines can be treated as independent entities. When applied to isolated vortices, the strong-pinning theory describes the competition between the energy gain provided by interaction with material defects and the energy cost associated with the deformation of the vortex line from its unperturbed straight configuration. Consider a vortex oriented along $z$ (crystallographic $c$-axis) and brought to rest upon decreasing the external current $j$ (applied along $y$) below a critical value $j_{c}$, as illustrated in the upper part of Fig.~\ref{fig:vortices_1d_3d}. In this dynamic pinning scenario, the typical longitudinal displacements $u$ between neighboring pins is determined by the pin-breaking condition
\begin{equation}
	\frac{\eone}{L} u \approx f_{p}.
	\label{eq:force-balance-equation-u-vs-f}
\end{equation}
where $\eone \approx \varepsilon_{0}/\gamma^{2}$ (up to logarithmic corrections) denotes the vortex line tension in an anisotropic system, with the uniaxial anisotropy parameter $\gamma$ and the vortex energy scale $\varepsilon_{0} = (\Phi_{0}/4\pi \lambda)^{2}$. Solving Eqs.~\eqref{eq:geom-relation} and \eqref{eq:force-balance-equation-u-vs-f} for $u$ and $L$, one arrives at
\begin{equation}
	u_{\sss \mathrm{1D}} \approx\Bigl(\frac{f_{p}}{n_{p}\up \eone}\Bigr)^{1/2}
	\quad \mathrm{and} \quad
	L_{\sss \mathrm{1D}} \approx \Bigl(\frac{\eone}{n_{p}\up f_{p}}\Bigr)^{1/2}.
	\label{eq:u-L-1D}
\end{equation}
Here, the subscript `$\mathrm{1D}$' indicates the limiting case of isolated vortices,%
\footnote{The term `1D strong-pinning theory' should not be confused with `1D pinning centers', where the latter denotes elongated defects that pin vortices over a large portion of their length.}%
i.e., where $B \to 0$. Substituting Eq.~\eqref{eq:u-L-1D} into Eq.~\eqref{eq:CritCurrTrapSegm}, we find the following expression for the critical current
\begin{equation}
	j_{c}^{\sss \mathrm{1D}} 
	\approx \frac{c f_{p}}{\Phi_{0}} (n_{p}\up)^{1/2} \Big(\frac{f_{p}}{\eone}\Big)^{1/2}.
	\label{eq:jc-1D}
\end{equation}
In the simplest case, one may assume the transverse trapping length $\up$ to be of the order of the defect's lateral diameter $a$, i.e., $\up \approx a$. This results in a critical current growing with the square-root of the defect density,  $j_{c} \propto n_{p}^{1/2}$, a result obtained earlier in Refs.~\cite{OvchinnikovI1991, BlatterGK2004}. By construction, the critical current is independent of the field strength~$B$. Langevin-dynamics simulations~\cite{KoshelevK2011} provide the following quantitative result
\begin{equation}
	j_{c} \approx 1.9 \frac{c}{\Phi_{0}}  \frac{\sqrt{n_{p} a}f_{p}^{3/2}}{\sqrt{\eone}}.
	\label{eq:IsolVortCritCurr}
\end{equation}
A more careful treatment~\cite{KoshelevK2011} suggests that the length $\up$ is determined by the distance at which the vortex undergoes a trapping instability. This instability depends on the pinning potential and yields the weak correction $\up \approx a (\varepsilon_{0}^{2} / \eone f_{p} n_{p} a^{3})^{1/9}$ for a single flux line.

\begin{figure}[tb]
	\centering
	\includegraphics[width=0.47\textwidth]{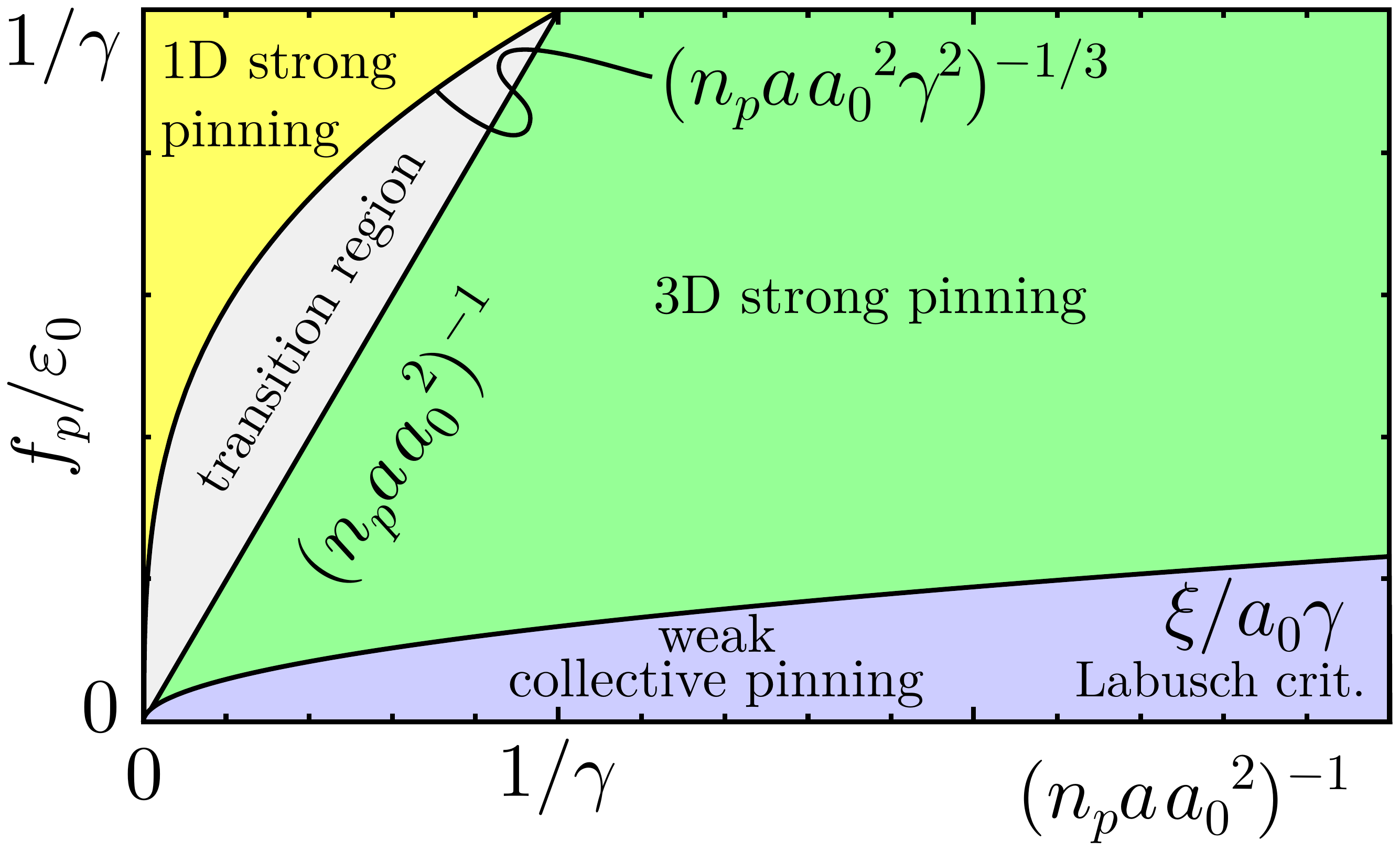}
	\caption{
		Strong-pinning regimes and their applicability boundaries, 
		see Eqs.~\eqref{eq:cond1D} and \eqref{eq:cond3D}, 
		for an anisotropic ($\gamma$) superconductor. The horizontal-axis 
		scale primarily features the dependence on field 
		$B \propto a_{0}^{-2}$ and defect concentration $n_{p}$, 
		while the vertical axis captures the dependence on the defect 
		strength $f_{p}$. At small pinning forces the prerequisite of strong 
		vortex pinning is not given and the system is described within 
		the theory of weak collective pinning. For illustrative purposes 
		we have neglected the weak field-dependence of $\up \approx a$.
	}
	\label{fig:phase_boundaries}
\end{figure}

The single-vortex regime holds until interactions between the vortices start to influence the pinned configuration. The typical vortex-vortex interaction force per unit length amounts to $\varepsilon_{0}/a_0$, where $a_0=(\Phi_0/B)^{1/2}$ is the intervortex spacing. This interaction can be treated as a small perturbation only if the force $\varepsilon_{0} L_{\sss \mathrm{1D}}/a_0$ acting on the pinned segment from other vortices  is smaller than $f_{p}$, giving the condition
\begin{equation}
	a_0 > \frac{\varepsilon_{0} \sqrt{\eone}}{f_p^{3/2} \sqrt{n_p \up}}.
	\label{eq:cond1D}
\end{equation}
An additional condition follows from the requirement that the pin-to-pin displacement $u_{\sss \mathrm{1D}}$ has to be smaller than the intervortex spacing $a_0$, yielding 
\begin{equation}
	a_0 > \frac{\sqrt{f_p}}{\sqrt{\eone n_p \up}}.
	\label{eq:cond1Du}
\end{equation}
For  $f_p < \sqrt{\eone \varepsilon_{0}}$, the last requirement is less restrictive than the previous one. Since  $\sqrt{\varepsilon_{0} \eone}$ defines an upper limit for the pin-breaking force $f_{p}$, Eq.~\eqref{eq:cond1Du} never limits the applicability of the 1D strong-pinning regime, meaning that this regime breaks down when the condition in Eq.~\eqref{eq:cond1D} is met, i.e., when $B/\Phi_{0} \approx (f_{p}^{3}/\varepsilon_{0}^{2} \eone) n_{p} \up$. A phase diagram marking the boundary line in Eq.~\eqref{eq:cond1D} is shown in Fig.~\ref{fig:phase_boundaries}. Other boundaries in this diagram will be discussed below.

\subsection{3D strong-pinning theory} \label{sec:3D}

At moderately high magnetic fields vortices form an ordered Abrikosov lattice, weakly deformed by separated material defects, see bottom of Fig.~\ref{fig:vortices_1d_3d}. This case is described by the theory of 3D strong pinning\footnote{The 3D strong-pinning theory assumes that isolated defects locally deform the vortex lattice without destroying its periodicity. For the case when the lattice is strongly deformed (or even destroyed) by the defects, no estimate for the critical current exists.} \cite{Labusch1969, LarkinO1979, OvchinnikovI1991}. Defects are assumed to be (i) sufficiently strong to produce a non-zero average pinning force while (ii) not yet strong enough to trap more than one flux line at a time. Consider a straight vortex line (along $z$) separated from the defect by $\vec{r} = (x,y)$. Its interaction with the defect deforms the flux line; a deformation that is uniquely characterized by its maximum value $\vec{u}$ at the height of the defect.  In the resulting planar problem, the deformation $\vec{u}$ generates an elastic restoring force $-\C \vec{u}$, where the effective spring constant $\C$ can be expressed through the elastic Green's function $G(\vec{r})$~\cite{BlatterGK2004} and includes contributions from both the vortex line tension and its interaction with the rest of the lattice, $\C \approx 3\sqrt{\eone \varepsilon_{0}}/a_{0} \approx (B / H_{c2})^{1/2} (\eone \varepsilon_{0} / \xi^{2})^{1/2}$, where $H_{c2} = \Phi_{0}/2\pi \xi^{2}$ is the upper critical field and $\xi$ is the coherence length. For a given (asymptotic) vortex position $\vec{r}$, the displacement $\vec{u}$ is determined by the balance condition between the restoring and pinning forces,
\begin{equation} 
	\C \vec{u}(\vec{r}) = \vec{f}_{p} \bigl[ \vec{r} + \vec{u}(\vec{r}) \bigr].
	\label{eq:non-linearfb}
\end{equation} 
The necessary ingredient for the existence of a finite average pinning force is that the function $\vec{u}(\vec{r})$ is multivalued in the range $\up < |\vec{r}|< u_{\sss \mathrm{3D}}$. Such multivalued region exists if the Labusch parameter~\cite{Labusch1969} $\kappa \equiv \max_{x}[f_{p}'(x)]/\C$ is larger than unity. Among the multiple solutions, the one that is realized $\vec{u}^{\mathrm{o}}(\vec{r})$ determines the pinning force $\vec{f}_{\!\mathrm{pin}}(\vec{r}) \equiv \vec{f}_{\!p}[\vec{r} + \vec{u}^{\mathrm{o}}(\vec{r})]$. Due to the appearance of multiple solutions in Eq.~\eqref{eq:non-linearfb}, this force function has jumps.

\begin{figure}[tb]
	\centering
	\includegraphics[width=0.35\textwidth]{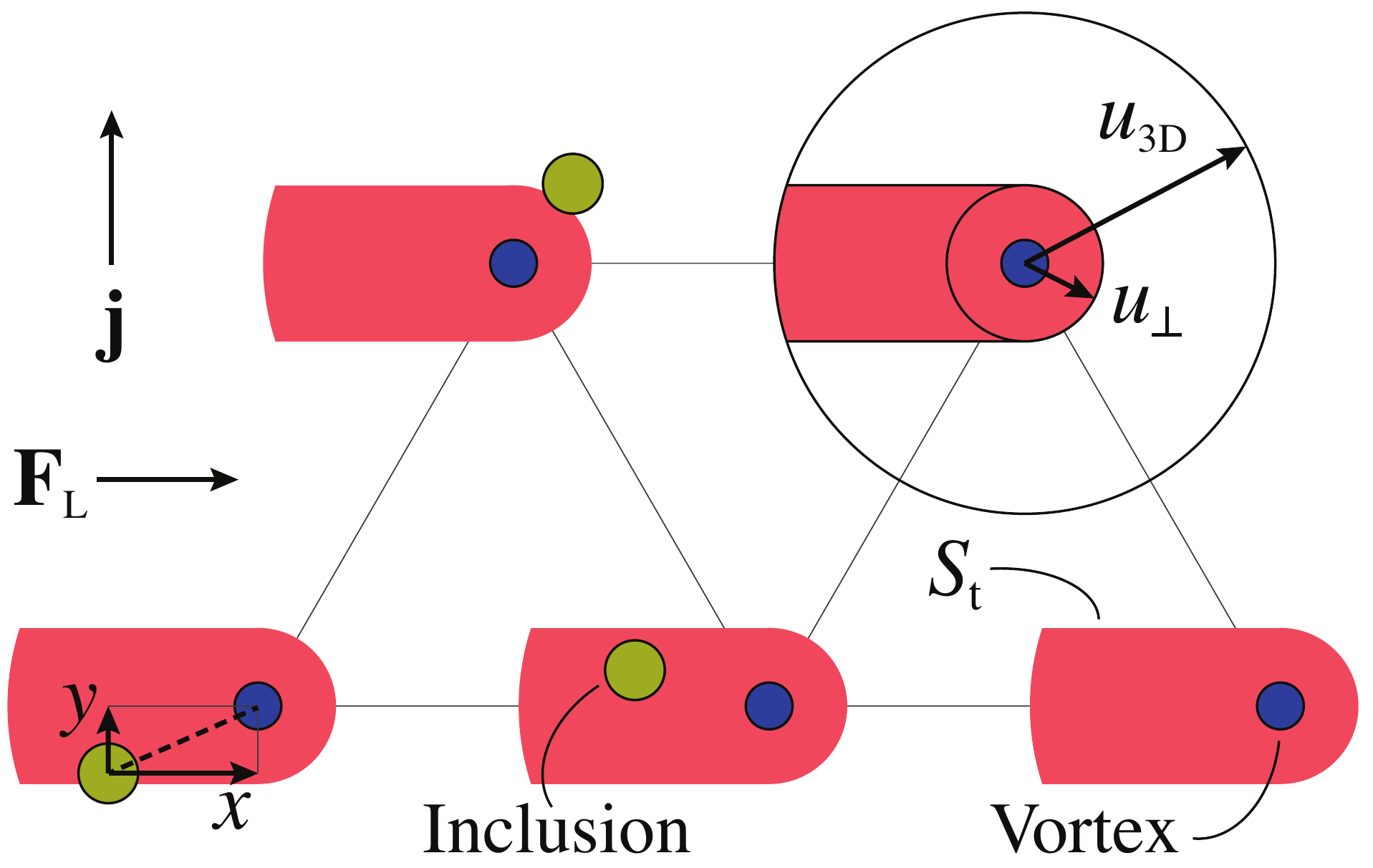}
	\caption{
		Illustration of trapping areas $S_{t}$ around the ideal lattice 
		positions at the depinning transition. In the critical state only 
		defects within these trapping areas capture vortex lines.
	}
	\label{fig:trap_area}
\end{figure}

In the dynamic scenario, a vortex line gets pinned when passing near a defect along $x$ at impact distance $y$ smaller than $\up$. The defect holds the vortex as long as the force $f_{\mathrm{pin}}(\vec{r})$ is smaller than the pin-breaking force $f_{p}$ providing the condition for the maximum possible deformation
\begin{equation}
	u_{\sss \mathrm{3D}} = f_{p}/\C.
	\label{eq:u-3D}
\end{equation}
As $u_{\sss \mathrm{3D}} > \up $,  the vortex lines are stronger deformed in the direction of motion, $|u_x|> |u_y|$. Only defects located within a so-called trapping area $S_{\mathrm{t}}$ capture vortex lines, see Fig.~\ref{fig:trap_area}. In the critical state, this area is defined by the conditions $|y|<\up$, $r< u_{\sss \mathrm{3D}}$ for $x>0$ and $r< \up$  for $x<0$. As a result, the fraction of occupied pins can be estimated as $\nu_{\mathrm{fill}} = S_{t}B / \Phi_{0} \approx u_{\sss \mathrm{3D}}\up/a_0^2$. Defects located outside the trapping area are empty, i.e., do not capture a vortex, and hence do not contribute to the bulk pinning force (density) $F_{c}$. The latter results from averaging $f_{\mathrm{pin},x}(\vec{r})$ over the trapping area, resulting in
\begin{equation}
	F_{c} = \frac{B}{c}j_{c}^{\sss \mathrm{3D}}
	= n_{p}\frac{B}{\Phi_{0}}\int_{S_{\mathrm{t}}}f_{\mathrm{pin},x}(\vec{r})d\vec{r}.
	\label{eq:Fp-3D}
\end{equation}
Deeply in the strong-pinning regime where $u_{\sss \mathrm{3D}}\gg \up$, the above integral simplifies and the critical current can be estimated as
\begin{equation} 
	j_{c}^{\sss \mathrm{3D}} \approx \frac{c}{\Phi_{0}} n_{p}f_{p}\up u_{\sss \mathrm{3D}} 
	\approx \frac{c}{\Phi_{0}} n_{p}\up \frac{f_{p}^2}{\C}.
	\label{eq:jc-3D}
\end{equation}
Alternatively, combining the geometric constraint, Eq.~\eqref{eq:geom-relation}, with the strong-pinning deformation, Eq.~\eqref{eq:u-3D}, one finds
\begin{equation}
	L_{\sss \mathrm{3D}} = \frac{\C}{n_{p} \up f_{p}}
	\label{eq:L-3D}
\end{equation}
for typical length of pinned segment, which---when inserted into Eq.~\eqref{eq:CritCurrTrapSegm}---provides the same estimate for $j_{c}^{\sss \mathrm{3D}}$ as that given in Eq.~\eqref{eq:jc-3D}. Let us highlight here that the critical current grows linearly with the defect density $n_{p}$ and decreases with the field strength as $B^{-1/2}$ (through $\C$). This scaling is again based on the simplest assumption that the transverse trapping length is determined by the defect diameter, $\up \approx a$. In reality, however, the situation is more complicated. A pinning potential typically decays as $-\mathcal{K} r^{-2}$ away from the defect. The coefficient $\mathcal{K}$ can be estimated as $\mathcal{K} \approx f_{p}\xi^{3}$ for small defects $a < \xi$  and as $\mathcal{K} = A\varepsilon_0 V_p$ for large (insulating) inclusions with $a > \xi$, where $V_p$ is the inclusion volume  and $A$ is the geometrical factor (for spherical inclusions in anisotropic superconductors $A \approx 2/\pi$). This long-range tail leads to a trapping instability and the field-dependent trapping distance $\up = 4[4\mathcal{K}/(27\C)]^{1/4} \approx [(\mathcal{K}^{2} \xi^{2}/\varepsilon_{0} \eone) (H_{c2}/B)]^{1/8}$. In that case, the critical current is expected to scale as $B^{-5/8}$~\cite{OvchinnikovI1991}.

Due to the confinement by neighboring vortices, the distortion $u$ imposed on the pinned vortex at the height of the defect decays along the flux line on a typical healing length $L_{h} = a_0\sqrt{\eone/\varepsilon_{0}}$. The 3D strong-pinning approach is justified when this healing length is shorter than the vortex segment length $L_{\sss \mathrm{3D}}$ providing the following criterion
\begin{equation}
	a_0 < \frac{\sqrt{\varepsilon_{0}} }{\sqrt{f_p n_p \up}}.
	\label{eq:cond3D}
\end{equation}
When this condition is violated, the vortex line wanders from one defect to the next without returning to its equilibrium position in the lattice and consequently, the defects do not act independently any more. This defines the boundary of the 3D pinning regime, see Fig.~\ref{fig:phase_boundaries}. For pin-breaking forces $f_p < \sqrt{\eone \varepsilon_{0} }$, the condition \eqref{eq:cond3D} differs from the break-down condition of the single-vortex (1D) regime, Eq.~\eqref{eq:cond1D}, suggesting the existence of an intermediate field range, see Fig.~\ref{fig:phase_boundaries}, defined by
\begin{equation}
	\frac{f_p^{3}}{\varepsilon_{0}^{2} \eone}n_p \up < \frac{B}{\Phi_{0}} < \frac{f_p}{\varepsilon_{0}} n_p \up,
	\label{eq:interm}
\end{equation}
where interactions between vortices are already essential but not yet strong enough to form an ordered lattice. These inequalities define the transition region shown in the phase diagram sketched in Fig.~\ref{fig:phase_boundaries}. Currently, no simple estimate for the critical current exists in this regime.

In order to fill this gap, we consider corrections to the 3D strong-pinning result, Eq.~\eqref{eq:jc-3D}, arising from events where multiple defects are found within the same healing volume $V_{h} = L_{h} u \up$. These events are rare when $n_{p}V_{h}$ is small. For the present analysis, we limit ourselves to those cases where two defects (a so-called doublet) share the same healing volume, an event that occurs with probability $(n_{p} V_{h})^{2}$. Any larger number $M > 2$ of multiplets occurs with a parametrically smaller probability $(n_{p}V_{h})^{M}$ and shall therefore be neglected here.

We expect a `typical' doublet to be stronger than one but weaker than two isolated defects, and hence the correction to the critical current, $\delta j_{c}=j_{c} - j_{c}^{\sss\mathrm{3D}}$, to be negative. On general grounds, we estimate this correction as $\delta j_{c} = -\eta_{d} n_p V_h j_{c}^{\sss\mathrm{3D}}$, with $\eta_{d}$ a positive number of order unity.

A more rigorous analysis requires averaging over different doublet realizations. In the following we derive a general framework to address this problem. Working in the reference frame of one defect, let the in-plane coordinate $\vec{r} = (x,y)$ define the distance to nearest (undeformed) vortex directed along $z$. With $\vec{R}_{s}=(X_{s}, Y_{s}, Z_{s})$ the position of the second defect, the in-plane distance of the latter to the vortex reads $\vec{r}-\vec{R}_{s}^{\perp} = (x-X_{s},\ y-Y_{s})$. While two defects---when considered isolated from each other---act on the vortex with forces $f_{\mathrm{pin}} (\vec{r})$ and $f_{\mathrm{pin}}(\vec{r}-\vec{R}_{s}^{\perp})$, the defect doublet will act with a force $f_{d}(\vec{r},\vec{R}_{s})$. Therefore this particular doublet leads to a correction of the total pinning force $F_{c} V$ [see Eq.~\eqref{eq:Fp-3D}] by $\delta f_{d}(\vec{r},\vec{R}_{s}) = f_{d}(\vec{r}, \vec{R}_{s}) - [f_{\mathrm{pin}} (\vec{r}) + f_{\mathrm{pin}}(\vec{r}-\vec{R}_{s}^{\perp})]$. Averaging over the two free coordinates $\vec{r}$ and $\vec{R}_{s}$, we obtain the correction to the critical current
\begin{multline}
	\delta j_{c}=\frac{2c}{\Phi_{0}}n_{p}^{2}\int d^{2}\vec{r}\int d^{3}\vec{R}_{s}  [f_{d}(\vec{r},\vec{R}_{s}) \\
	- f_{\mathrm{pin}}(\vec{r})-f_{\mathrm{pin}}(\vec{r}-\vec{R}_{s}^{\perp})].
	\label{eq:Corrjc}
\end{multline}
The evaluation of this double integral within an elastic model, see Appendix~\ref{sec:doublet}, yields the quantitative estimate
\begin{equation}
	\delta j_{c} \approx - \frac{2}{3} j_{c}^{\sss \mathrm{3D}} n_{p}V_{h},
	\label{eq:doublet-correction}
\end{equation}
following our expectation with $\eta_{d} = 2/3$. As $V_h \propto B^{-1}$, this correction scales with field
roughly as $B^{-3/2}$ and becomes important at lower fields.

\subsection{High fields: full-occupation regime} \label{sec:high_fields}

At large fields $B > B_{\mathrm{hf}}$, each inclusion---independently of its position---captures a vortex line. The criterion
\begin{equation}
	a_{0} = 2\up,
   	\label{eq:bare-hf-criterion}
\end{equation}
translates into a crude estimate for $B_{\mathrm{hf}} \approx \Phi_{0}/4\up^{2}$, marking the break-down of the 3D strong-pinning theory. In Fig.~\ref{fig:trap_area}, the entire area is now covered in red. Since all particles are occupied by (at least) one vortex, the critical current assumes the simplified form
\begin{equation}
	j_{c}^{\mathrm{hf}} = \beta \frac{c f_{p}}{B} n_{p},
	\label{eq:jc-hf}
\end{equation}
where $\beta < 1$ is a numerical factor appearing due to averaging over the pin positions. If the field dependence of $f_{p}$ is weak, $j_{c}$ decays inversely proportional to the field strength $B$ while growing linearly with the defect density $n_{p}$. This scaling is also obtained from the 3D strong-pinning result, Eq.~\eqref{eq:jc-3D}, after substituting both longitudinal and transverse trapping lengths by the intervortex distance, i.e., $u = \up \approx a_{0}/2$.

It occurs, however, that the high-field regime cannot be characterized by the simple $1/B$ law suggested Eq.~\eqref{eq:jc-hf}. In fact, when the intervortex distance $a_{0}$ becomes comparable to the full longitudinal length of the trapping area $u_{\sss \mathrm{3D}} + \up$, the pin-breaking force acquires a significant field dependence. Indeed, once the nearest unpinned vortex approaches the defect to a distance comparable to $\up$, it may undergo a pinning instability even if the defect is already occupied. This instability can be quantified by studying a set of coupled force-balance equations similar to Eq.~\eqref{eq:non-linearfb} for two neighboring vortices  (see Appendix~\ref{sec:double_occupation} for more details). At the second trapping instability, the already pinned vortex leaves the defect due to the arrival of the newly pinned flux line and before reaching its 'bare' critical state. The quantitative criterion for the appearance of the instability-limited critical state can be expressed through
\begin{equation}
	a_{0} = \up + (1-\Gamma)u_{\sss \mathrm{3D}}
\end{equation}
and is derived in Appendix~\ref{sec:double_occupation}. Here, $\Gamma = G(a_{0})^{-1}/G(0)^{-1}$ denotes an elastic coupling coefficient. A quantitative analysis provides us with the estimate $\Gamma \approx 0.23$. Neglecting the weak field-dependence of $\up \approx a$, this instability arises when
\begin{equation}
	a_{0} = \frac{a}{1- (1-\Gamma)f_{p}/3\sqrt{\varepsilon_{0} \eone}} = \beta_{c} a,
\end{equation}
with $\beta_{c} > 1$. Simulations, discussed below, suggest that $2 < \beta_{c} < 3$. Although distinct, the closeness of this instability to the criterion \eqref{eq:bare-hf-criterion} makes it technically difficult to separate these two transitions. Most prominently, this phenomenon will lead to a decreasing $f_{p}(B)$ (upon increasing $B$) and hence the critical current will decay faster than $B^{-1}$.

Another, yet more spectacular effect occurs when the defect traps two (or even more) vortices, i.e., when the pinning instability of the second vortex is not accompanied by the departure of the first one. At this moment the pin-breaking force $f_{p}(B)$ experiences a strong revival leading to a novel type of peak effect. This case is briefly discussed in Sec.~\ref{sec:single_inclusion} below and appears (empirically) when the intervortex distance falls below $2a$, i.e., for $B > \Phi_{0}/4a^{2}$.

\section{Time-dependent Ginzburg-Landau model for numerical simulations} \label{sec:model}

The numerical results presented in this paper are obtained using an iterative, massive-parallel solver for the time-dependent Ginzburg-Landau (TDGL) equation suitable for large three-dimensional systems with typical sizes of $100$ coherence lengths in all three directions. The technical details of the numerical algorithm and a benchmark analysis for its implementation on graphics procession units (GPUs) are described in Ref.~\cite{SadovskyyJComp2015}. Here we only present the used dimensionless form and notations of the TDGL equations. The dynamics of the superconducting order parameter $\psi(\vec{r},t)$ is described by the TDGL equation
\begin{multline}
	\mathfrak{u}(\partial_t + i \mu)\psi = \epsilon(\vec{r})\psi- |\psi|^2\psi \\ 
	+ \sum_{k=x,y,z} \tilde{\xi}_k^2(\nabla_k - i A_k)^2\psi + \zeta(\vec{r},t).
	\label{eq:TDGL}
\end{multline}
Here, all lengths are measured in units of the in-plane coherence length $\xi$ at the simulated temperature, such that $\tilde{\xi}_x = \tilde{\xi}_y = 1$ and $\tilde{\xi}_z = 1/\gamma$, with $\gamma$ being the uniaxial anisotropy factor. The time $t$ is measured in units of $t_{0} = 4\pi \sigma_{n} \lambda^{2} / c^{2}$, where $\sigma_{n} = 1/\rho_{n}$ is the normal state conductivity, $\lambda$ the in-plane penetration depth, and $c$ the speed of light. The function $\epsilon(\vec{r})$ captures the local critical temperature of the sample. By changing its value from unity in the bulk\footnote{In Ref.~\cite{SadovskyyJComp2015} a slightly different normalization is used, where $\epsilon$ has the value $T_c / T-1$ in the bulk and the unit of length is the zero-temperature coherence length $\xi(0)$. It is straightforward to show that this choice is equivalent to fixing $\epsilon = 1$ in the bulk, while normalizing all lengths to the coherence length $\xi(T) = \xi(0) / \sqrt{T_c/T-1}$.} to $\epsilon(\vec{r})=-1$ in specific regions, we can model normal inclusions. We use the infinite-$\lambda$ approximation which describes superconductors at high magnetic fields when the penetration depth $\lambda$ is much larger than the distance between the vortex lines $a_0=\sqrt{\Phi_0/B}$. In this approximation the vector potential is fixed by the external field. In the Landau gauge, the dimensionless vector potential takes the form $\vec{A} = [0, (B_{z}/H_{c2}) x ,0]$, for an external magnetic field applied along the $c$-axis, and $H_{c2}=\Phi_{0}/(2\pi\xi^2)$ being the corresponding upper critical field.

The system's temporal evolution depends on the reduced relaxation rate $\mathfrak{u}$ and the scalar electric potential $\mu$, while thermal noise is accounted for by the $\delta$-correlated Langevin term $\zeta (\vec{r},t)$,
\begin{equation}
	\langle\zeta^*(\vec{r},t) \zeta(\vec{r}',t') \rangle
	= \mathfrak{u} T \, \delta(\vec{r} - \vec{r}' ) \delta(t - t').
	\label{eq:noise_zeta}
\end{equation}
In the above expression, $T$ is the reduced temperature measured in units of $H_c^2\xi^3/8\pi$, with $H_c = \Phi_{0} / 2\sqrt{2} \pi \lambda \xi$ the thermodynamic critical field. In a generic simulation setting, the magnetic field is aligned along the $z$-axis (or $c$-axis) and the current applied along the $y$ direction (full-force configuration). The electric current is measured in units of $j_0=2c\varepsilon_0/(\Phi_0\xi) $ (cgs), which gives for the depairing current $j_{\mathrm{dp}}=(2/3\sqrt{3}) j_{0} \approx 0.385 j_{0}$. The total dimensionless current along $y$ reads
\begin{equation}
	j = \mathrm{Im} [ \psi^*(\nabla_{\!y} -i A_y)\psi ] 
		- \partial_t A_y - \nabla_{\!y} \mu,
	\label{eq:Jgen}
\end{equation}
where the first term describes the supercurrent and the normal current is given by the last two terms. The dimensionless electric field (along $y$) $E = -\partial_t A_y - \nabla_y \mu$, generated by the flux motion, is measured in units of $E_{0}=\xi H_{c2}/ct_{0}$. For the simulations discussed here, we used periodic boundary conditions in $x$ and $y$ direction, while the system had open boundaries along $z$. The implementation of a fixed current in the case of periodic boundary conditions is discussed in Ref.~\cite{SadovskyyJComp2015}.

\section{Pin-breaking force from an isolated inclusion} \label{sec:single_inclusion}

\begin{figure}[tb]
	\centering
	\subfloat{\includegraphics[width=0.45\textwidth]{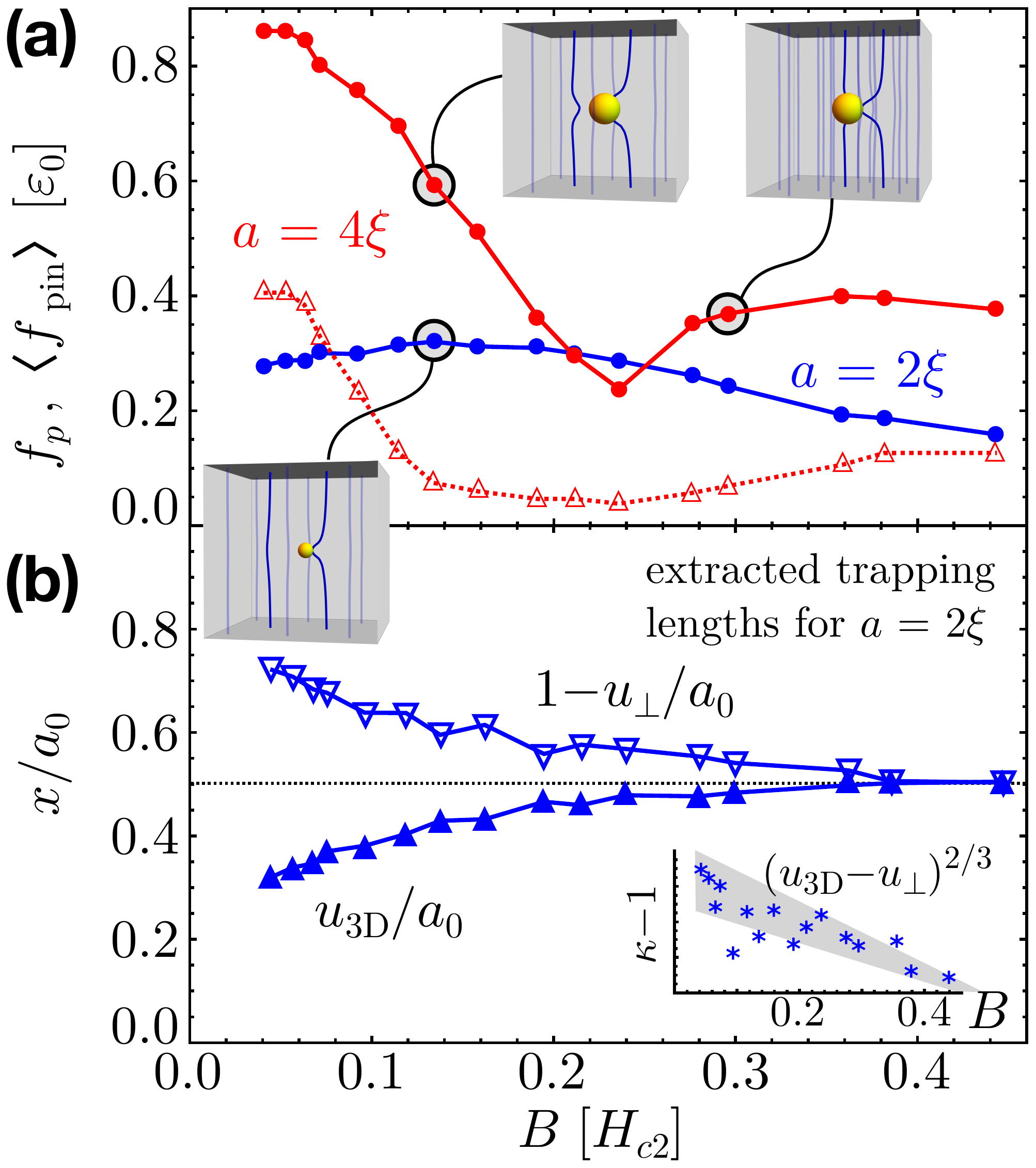}\label{fig:pin_breaking_force_a}}
	\subfloat{\label{fig:pin_breaking_force_b}}
	\caption{
		\protect\subref{fig:pin_breaking_force_a}~Solid symbols show 
		the pin-breaking force $f_{p}(B)$ of an isolated defect 
		for two different defect sizes $a = 2\xi$ (blue circles) and 
		$a = 4\xi$ (red circles). In both cases this force has a pronounced 
		and nonmonotonic field dependence. The kink in the $a = 4\xi$ 
		curve at $B = 0.23 H_{c2}$ is caused by the first-order 
		phase transition separating the singe- and double-occupied 
		ground states of an isolated inclusion. For the larger particle 
		with  $a = 4\xi$, open triangles show the pinning force 
		$\langle f_{\mathrm{pin}}(x)\rangle$ \textit{averaged} over 
		the lattice coordinate $x$. The insets illustrate examples 
		of the critical states where $f_{p}$ is realized.
		\protect\subref{fig:pin_breaking_force_b}~Coordinate 
		$x = u_{\sss \mathrm{3D}}$  ($x = a_{0} - \up$) at which 
		the pinned vortex leaves (the unpinned vortex is captured by) 
		the inclusion $a = 2\xi$, normalized to the intervortex 
		separation $a_{0}$. The values for $u_{\sss\mathrm{3D}}$ 
		and $\up$ being close suggests that the system approaches 
		the Labusch point $\kappa = 1$ where the defect looses its 
		`strong' property (see main text). Near this point the Labusch 
		parameter $\kappa$ is expected to scale as 
		$\kappa - 1 \propto (u_{\sss \mathrm{3D}} - \up)^{2/3}$, 
		see inset.
	}
	\label{fig:pin_breaking_force}
\end{figure}

The key quantity characterizing a defect's pinning capability is its pin-breaking force $f_p$, i.e., the maximal force with which an isolated inclusion can act on the vortex system. In order to facilitate a quantitative comparison between theory and simulations, we directly compute this parameter at different fields for the two particle sizes studied in this paper, i.e., $a = 2\xi$ and $4\xi$.  A detailed investigation of pinning properties of isolated inclusions will be published elsewhere. 

Figure~\ref{fig:pin_breaking_force} shows the magnetic-field dependence of the pin-breaking force $f_{p}$ for an isolated inclusion inside an ideal vortex lattice for two diameters. Simulation were done with 36 vortex lines by adjusting the system sizes $L_x$ and $L_{y}$ so that the $6 \times 6$ vortex lattice ideally fits into the system. Traditionally, it is assumed that $f_{p}$ is an intrinsic property of the defect, and hence independent of the field strength $B$. We observe, however, that $f_{p}$ does have a substantial field dependence, especially for  $a = 4\xi$. Moreover, this dependence is nonmonotonic. Several effects cause variation of $f_{p}$ with the magnetic field. For a single vortex, the pin-breaking force $f_{p}(0)$ is reached when the two branches of the pinned vortex tip form a critical angle. At small fields, $a_0 = (\Phi_{0}/B)^{1/2} \gg \up$, neighboring vortices will rectify the pinned vortex and enhance the angle between the tips meaning that the critical angle is reached at higher currents. As a consequence, at low fields the pin-breaking force \textit{increases} with increasing $B$, as observed for $a = 2\xi$. At intermediate fields, $a_0 \gtrsim \up$, the vortex approaching the defect along the force direction will compete with the pinned vortex and \textit{reduce} the pin-breaking force of the latter, see Appendix~\ref{sec:double_occupation} for a quantitative criterion. At sufficiently high fields, empirically for $a_0 \approx 2a$, the inclusion will accommodate two pinned vortices. The transition from the single-occupied to double-occupied ground state for $a = 4\xi$ can be seen as a kink in the $f_p(B)$ curve at $B \approx 0.23H_{c2}$. It is remarkable that at the kink $f_p$ drops below the pin-breaking force for $a = 2\xi$. Above this point, the  $f_p(B)$ sharply increases again. Pushing to even higher fields, when the competition with more unpinned vortices becomes relevant, the pin-breaking force will eventually decrease again. Further revivals of the pin-breaking force are observed each time the defect pins one more vortex (inclusions with diameter $a \geqslant 5\xi$, not shown here).

It is important to note that randomly distributed defects \textit{do not} act with the upper bound force $n_{p} f_{p}$ on the vortex system. Instead, each defect realizes a certain force $f_{\mathrm{pin}}(\vec{r})$, which is determined by the smallest pin-to-defect vector $\vec{r}$. Therefore, the maximal pinning force $n_{p} \langle f_{\mathrm{pin}}(\vec{r})\rangle$ results from proper averaging over all realized states. The simulations presented here allow us to calculate the average force $\langle f_{\mathrm{pin}}(x)\rangle \equiv (1/a_{0}) \int_{0}^{a_{0}} dx \, f_{\mathrm{pin}}(x)$ for a specific impact parameter $y = 0$. Indeed, for a system in the quasistatic regime, $j/j_{c} - 1 \ll 1$, we can rewrite the dynamic equation $f_{\mathrm{pin}} = N_{v} L_{z} (\eta v - \Phi_{0} j / c)$ in the form
\begin{equation}
	f_{\mathrm{pin}}[x(t)]/\varepsilon_{0} 
	= 2 N_{v} L_{z} [E(t) / \rho_\mathrm{ff} - j]/j_{0},
\end{equation}
where, $\eta \approx \Phi_0 H_{c2} / \rho_n c^2$ denotes the single-vortex viscosity and $\rho_n$ the normal-state resistivity. The relation between the viscous force $\eta v$ and electric field $E$ associated with the vortex motion is obtained from independent simulations of a defect-free system, as reported in Ref.~\cite{KoshelevPRB16}. In this case one has $\eta v = \Phi_{0}j/c = \Phi_{0} E / \rho_\mathrm{ff} c$ and the flux-flow resistivity $\rho_\mathrm{ff}$ has been numerically evaluated as  $\rho_\mathrm{ff} = 1.689 (B/H_{c2}) \rho_{n}$. All ingredients necessary to evaluate the above expression, i.e., the coordinate $x$, the electric field $E$ and the applied current $j$, can be extracted at given simulation times\footnote{The extraction of $x$ requires an analysis of the order parameter state at a given time $t$. Due to limitations in the numerical capacity of generating/analyzing this output for each simulation time step $t$, we typically limit ourselves to times $t_{i}$ separated by $\delta N_{t} = 10^{4}$ simulation steps ($\delta N_{t}/N_{t} \approx 5 \times 10^{-3}$). In order to reduce numerical noise, we further average $E(t)$ over this time-window.} $t$. The averaged pinning force for $a = 4\xi$ extracted in this way is plotted in Fig.~\subref*{fig:pin_breaking_force_a}. We observe that its behavior is different from the maximum pinning force; $\langle f_{\mathrm{pin}}(x)\rangle$ does not have maximum at small fields and its minimum near the double-occupation transition is rather shallow. While deep in the strong-pinning limit, the theory of strong vortex pinning predicts $\langle f_{\mathrm{pin}}(\vec{r})\rangle / f_{p} \approx u_{\mathrm{3D}}\up/a_{0}^{2} \propto B^{1/2}$, simulations for $a$ from $2\xi$ to $4\xi$ suggest that the ratio $\langle f_{\mathrm{pin}}(\vec{r})\rangle/f_{p}$ is in the range $[1/9, 1/3]$, lacking a simple field-dependence due to the non-monotonicity of $f_{p}(B)$.

\begin{figure*}[tb]
	\centering
	\includegraphics[width=\textwidth]{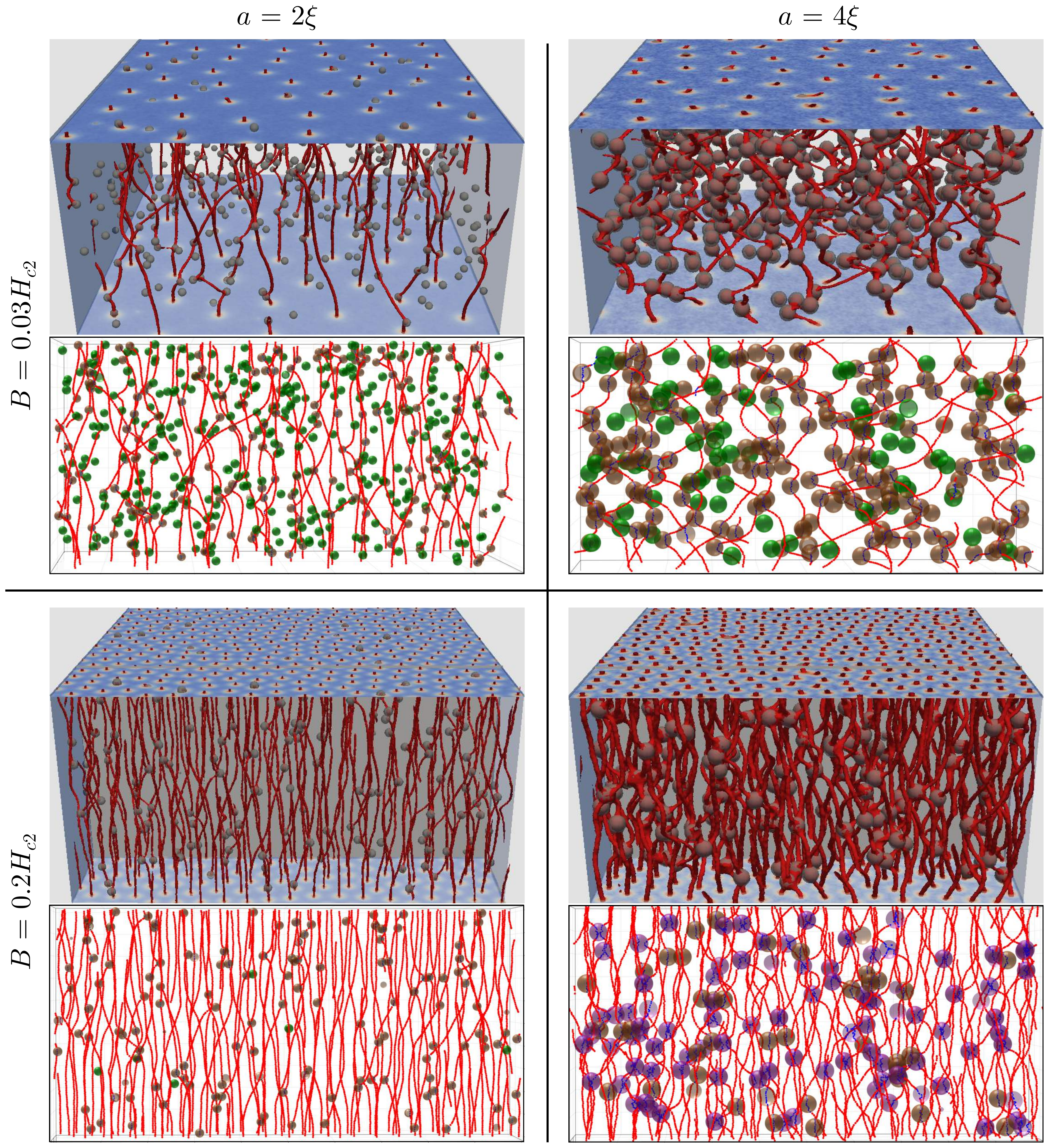}
	\caption{
		Each quadrant represents a pinned vortex configurations 
		for either of two fields $B = 0.03H_{c2}$ and $0.2H_{c2}$ 
		and for 500 defects ($n_p\xi^3=10^{-3}$) either of size 
		$a = 2\xi$ or $a = 4\xi$. Within each quadrant, the top view 
		shows isosurfaces of the superconducting order parameter, 
		visualizing its suppression near vortices and within 
		the inclusions (semi-transparent spheres). The lower view 
		shows the same configuration after numerical extraction 
		of the vortex lines and analysis of their configuration. 
		Vortex segments outside (inside) inclusions are red (blue). 
		The color of the spheres codes the occupation state of 
		the particles: green for empty, brown for single occupied, 
		and purple for double occupied. See also supplementary 
		movie clips at applied current slightly above the critical 
		current and at magnetic field $B = 0.03 H_{c2}$ for 
		\uhref{https://youtu.be/aV4MZzUPYxs}{$a = 2\xi$} 
		(cf. \uhref{https://youtu.be/P07FsRceVbQ}{clip} at larger applied 
		current) and \uhref{https://youtu.be/nTyTnmLOs5o}{$a = 4\xi$}
		as well as at $B = 0.22 H_{c2}$ for 
		\uhref{https://youtu.be/ae0oME77Pz0}{$a = 2\xi$} and 
		\uhref{https://youtu.be/YEFLqfXslQs}{$a = 4\xi$}.
	}
	\label{fig:vortex_configs} 
\end{figure*}

Whereas this procedure works for large inclusions $a = 4\xi$, it does not provide reliable output for small defects $a = 2\xi$. In the latter case the average pinning force turns out to be close to zero. In order to analyze this situation further, we have extracted the vortex lattice's center-of-mass coordinates at which the pinned vortex line leaves the inclusion, $x = u_{\sss\mathrm{3D}}$, and at which the next vortex is captured again, $x = a_{0} - \up$. The obtained values, shown in Fig.~\subref*{fig:pin_breaking_force_b}, suggest that the small inclusion transits from strong ($u_{\sss\mathrm{3D}} > \up$) pinning to weak ($u_{\sss\mathrm{3D}} = \up$) pinning at high fields $B\approx 0.4 H_{c2}$. This transition has been predicted~\cite{Willa2016} for metallic defects in the vicinity of $H_{c2}$. Near the transition to weak pinning~\cite{Koopmann2004}, the (Labusch) parameter $\kappa \geqslant 1$ relates to the pinning lengths via $\kappa - 1 \propto (u_{\sss\mathrm{3D}} - \up)^{2/3} \ll 1$, see inset of Fig.~\subref*{fig:pin_breaking_force_b}, and the critical current (in its simplest form) is expected~\cite{BlatterGK2004, Koopmann2004} to scale as $j_{c} \propto (\kappa-1)^{2}$. Approaching the Labusch point $\kappa = 1$, may therefore have a much stronger effect on the critical current than the field dependence of $f_{p}(B)$.

\section{Pinning regimes and magnetic field dependences of the critical currents} \label{sec:field_dependence}

\begin{figure*}[tb]
	\centering
	\subfloat{\includegraphics[width=0.99\textwidth]{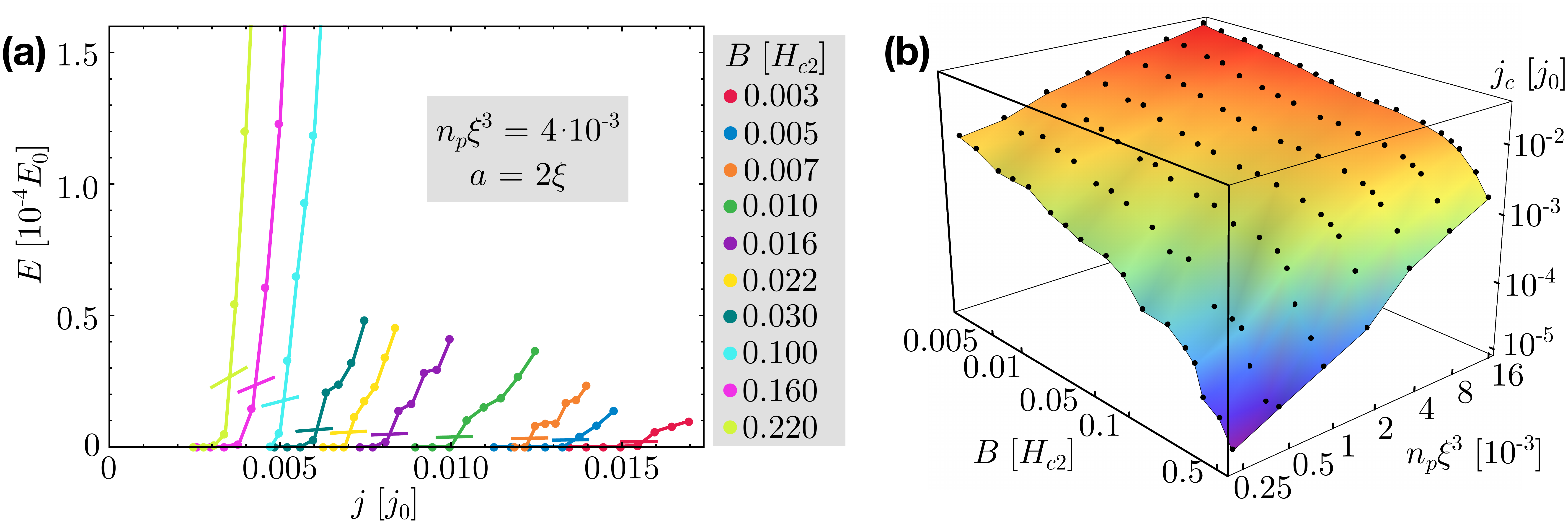}\label{fig:a2_iv_jc_a}}
	\subfloat{\label{fig:a2_iv_jc_b}}
	\caption{
		\protect\subref{fig:a2_iv_jc_a}~Representative set of current-voltage 
		dependences at fixed density $n_{p} = 4 \times 10^{-3}\xi^{-3}$ 
		of small defects ($a = 2\xi$) for different magnetic fields $B$. 
		Short lines show electric fields corresponding to 2\% of 
		the free flux-flow voltage. The intersection with the \IV curve 
		is used as a criterion for $j_{c}$.
		\protect\subref{fig:a2_iv_jc_b}~Surface plot of the critical current 
		$j_{c}$ for defects of size $a = 2\xi$ as a function of their density 
		$n_{p}$ and the magnetic field $B$. Black points indicate the 
		$j_{c}$-values used to create the map. Note the logarithmic scale 
		on all three axes. The thick frame indicates the plane of projection 
		in Fig.~\protect\subref*{fig:a2_jc_B_a}.
	}
	\label{fig:a2_iv_jc}
\end{figure*}

We systematically explored the evolution of the current-voltage (\IV) dependences for different magnetic fields, particle sizes, and particle densities. All numerical results presented below are obtained for a system of size $V = 100\xi \times 100\xi \times 50\xi$ with $256 \times 256 \times 128$ mesh points. The pinning landscape is modeled as a random distribution of $N_{p}$ of identical (metallic) spherical inclusions with diameter~$a$ and $\epsilon = -1$ inside [see Eq.~\eqref{eq:TDGL}]. The defect density $n_{p} = N_{p}/V$ or the `nominal' defect volume fraction $\nu_{\mathrm{vol}}^{0} = (\pi/6)n_{p}  a^{3}$ are independent of the system size and therefore more appropriate than $N_{p}$ to characterize the defect landscape. Notice that due to partially overlapping defects (which is noticeable  for $\nu_{\mathrm{vol}}^{0} \gtrsim 0.2$), the true volume fraction occupied by inclusions $\nu_{\mathrm{vol}}$ is somewhat smaller and well described by the expression $\nu_{\mathrm{vol}} \approx \nu_{\mathrm{vol}}^{0} - (\nu_{\mathrm{vol}}^{0})^{2}/2$. A selection of pinned vortex configurations near criticality is illustrated in Fig.~\ref{fig:vortex_configs}.

A typical simulation run consists of two phases. The system is initialized with (i)~a random order parameter, (ii)~an external current $j > j_{c}$, and (iii)~a relatively high thermal noise level. In a first phase the Langevin noise is slowly reduced and the system condenses into a dynamic vortex state moving over the pinning landscape. In a second phase, the noise level is kept small and the system is ramped through decreasing current values. At each new current value, the system is given time to find a `steady state' (typically $N_{t} = 5 \times 10^{5}$ time iterations) after which the electric field (or voltage) across the sample is recorded for the same time ($N_{t} = 5 \times 10^{5}$). Each pair of current~$j$ and averaged electric field $\langle E \rangle_{N_{t}}$ then represents one data point of the current-voltage characteristic. Typically, we did not observe significant history effects, i.e., {\IV} curves differing by the starting current and/or the current step size are close. Only at smallest magnetic fields/smallest defect densities {\IV} dependences become noisy and slightly history-dependent. In the case of low fields $B < 0.01 H_{c2}$, the reason lies in the insufficient number of vortex lines $N_{v} < 16$ to form a lattice. In the case of low defect densities, the critical current gets small, and the flux-line motion within the simulation time drops below a few coherence lengths $\xi$ leading to an ill-defined temporal averaging. We extract the critical current $j_{c}$ from the {\IV} curve using as a criterion the intersection of current-voltage characteristic with 2\% of the free flux-flow electric field $E(j_{c}) = 0.02 \rho_\mathrm{ff} j_c$.

In order to deepen our understanding of the pinning mechanisms, we have extracted the vortex lines from the order-parameter distributions using algorithm from Ref.~\cite{PhillipsPRE2015}, filtered out only field-induced vortices, and performed a detailed analysis of trapped vortex configurations. Typical snapshots of these configurations are illustrated in Fig.~\ref{fig:vortex_configs}. Once extracted, the vortex lines are split into line segments localized inside the metallic inclusions and threading superconducting regions. We then used this information to compute several parameters characterizing trapped configurations. These are (i) the fraction $\nu_{\mathrm{fill}}$ of particles occupied by vortices, (ii) the fractions $\nu_{>n}$ of particles occupied by more than $n$ vortices ($\nu_{>0}\equiv\nu_{\mathrm{fill}}$), (iii) the average length $L$ of trapped segments, and (iv) the average particle-to-particle displacements $u$ ($\up$) along (transverse to) the direction of vortex motion. The definitions of $L$, $u$, and $\up$ are illustrated in the upper right picture of Fig.~\ref{fig:vortices_1d_3d}.

\subsection{Small-size particles: comparison with strong-pinning theory} \label{sec:a2xi}

\begin{figure}[tb]
	\centering
	\includegraphics[width=0.47\textwidth]{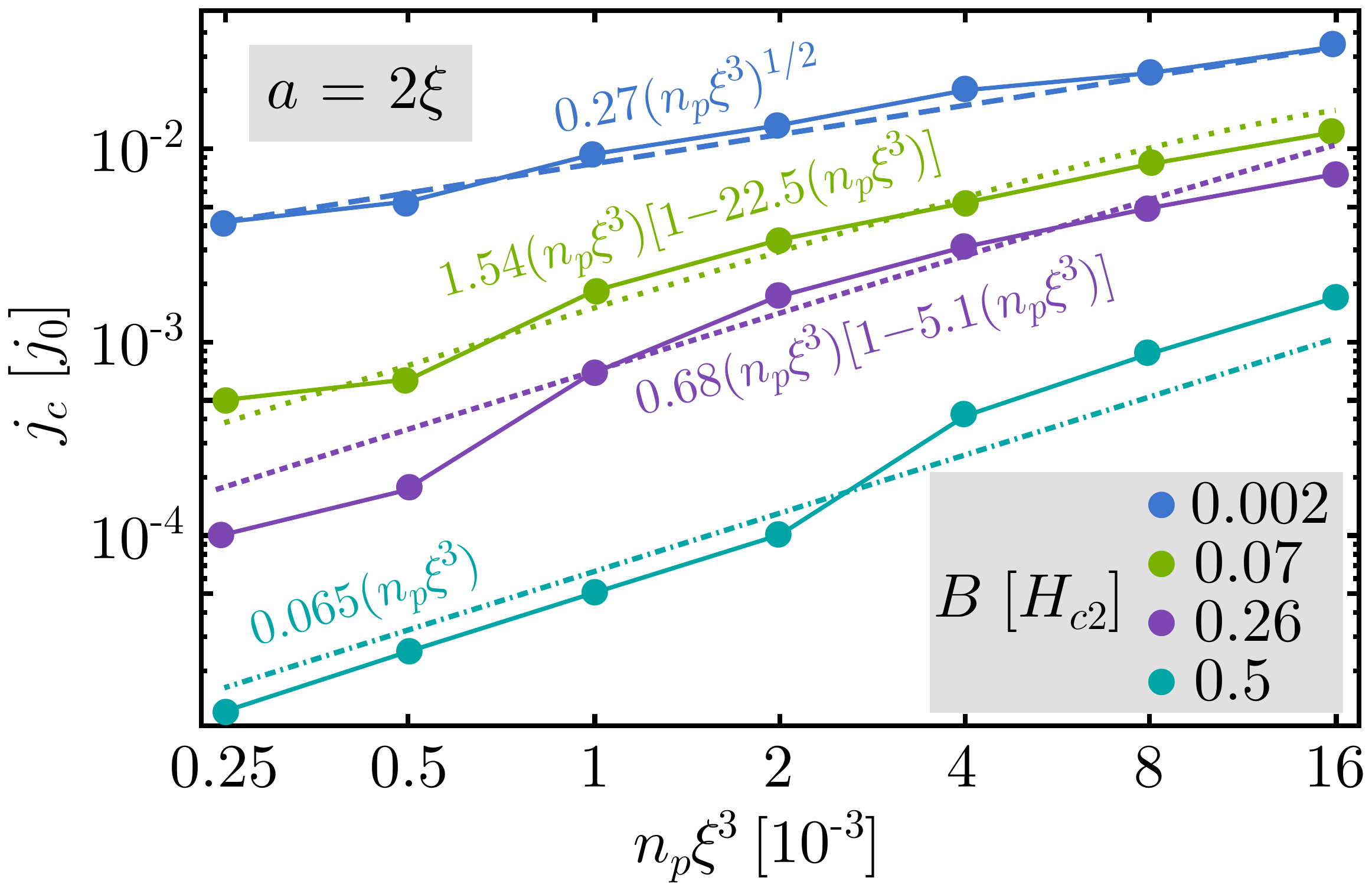}
	\caption{
		The inclusion-density dependences of the critical current 
		for selected magnetic fields. Lines show predictions from 
		strong-pinning theory for (i)~the 1D case (dashed) 
		at low fields, (ii)~the 3D case (dotted) at intermediate 
		fields, and (iii)~the high-field case (dash-dotted) when 
		all defects are occupied. The vortex lattice's order-disorder 
		transition is accompanied by a jumplike increase in $j_{c}$, 
		see also Fig.~\protect\subref*{fig:a2_jc_B_b}.
	}
	\label{fig:a2_jc_np}
\end{figure}

In this section we present results for small spherical particles with diameter $a = 2\xi$. A representative set of current-voltage characteristics used to determine $j_{c}$, is shown in Fig.~\subref*{fig:a2_iv_jc_a} for $n_{p}\xi^{3} = 4 \times 10^{-3}$. Performing simulations for multiple field values in the range $2 \times 10^{-3} \leqslant B/H_{c2} \leqslant 0.5$ and for a wide range of defect densities, $0.25 \times 10^{-3} \leqslant n_{p}\xi^{3} \leqslant 16 \times 10^{-3}$ (corresponding to volume fraction $0.001 \leqslant \nu_{\mathrm{vol}} \leqslant 0.064$), we have mapped out the critical current as a function of these two parameters, see Fig.~\subref*{fig:a2_iv_jc_b}. A visual impression of vortex arrangements in the critical state is given in Fig.~\ref{fig:vortex_configs} (left column).

Figure~\ref{fig:a2_jc_np} shows the dependence of the critical current on the defect density for four magnetic fields representing different scalings regimes of $j_c(n_p)$. At low fields, $j_{c}$ grows as $n_{p}^{1/2}$ (dashed line) as expected from 1D strong-pinning theory, see Eq.~\eqref{eq:jc-1D}. At intermediate fields, the growth is linear in $n_{p}$ with a weak downwards correction at larger densities. This effect is well captured by the 3D strong-pinning result, Eq.~\eqref{eq:jc-3D}, including the doublet contribution from Eq.~\eqref{eq:doublet-correction}, which we rewrite in a form convenient for comparison with simulations, 
\begin{equation} \label{eq:jc-comp-with-theory}
	\frac{j_{c}}{j_{0}} = \eta_{0} \gamma n_{p} \xi \up a_{0} \frac{f_{p}^2}{\varepsilon_{0}^2}
	\Big( 1 - \frac{\eta_{d}}{3} n_{p}\up a_{0}^2 \frac{f_{p}}{\varepsilon_{0}} \Big),
\end{equation}
where $\eta_{0}$ and $\eta_{d}$ are the numerical constants. Our simulations agree best with the theory when using $\eta_{0} \approx 1/6$ and a numerical coefficient for the doublet correction $\eta_{d}\approx 1$, close to the value 2/3 evaluated using a simple model in Appendix~\ref{sec:doublet}. Here we used $\up = (\gamma f_{p} a_{0} / 3\varepsilon_{0} a)^{1/4} a$ for the trapping instability length. We also observe that for intermediate fields at largest defect densities the critical current approximately grows again as $\sqrt{n_p}$. At the largest fields, the critical current grows linearly over the entire range of defect densities.

\begin{figure}
	\centering
	\subfloat{\includegraphics[width=0.5\textwidth]{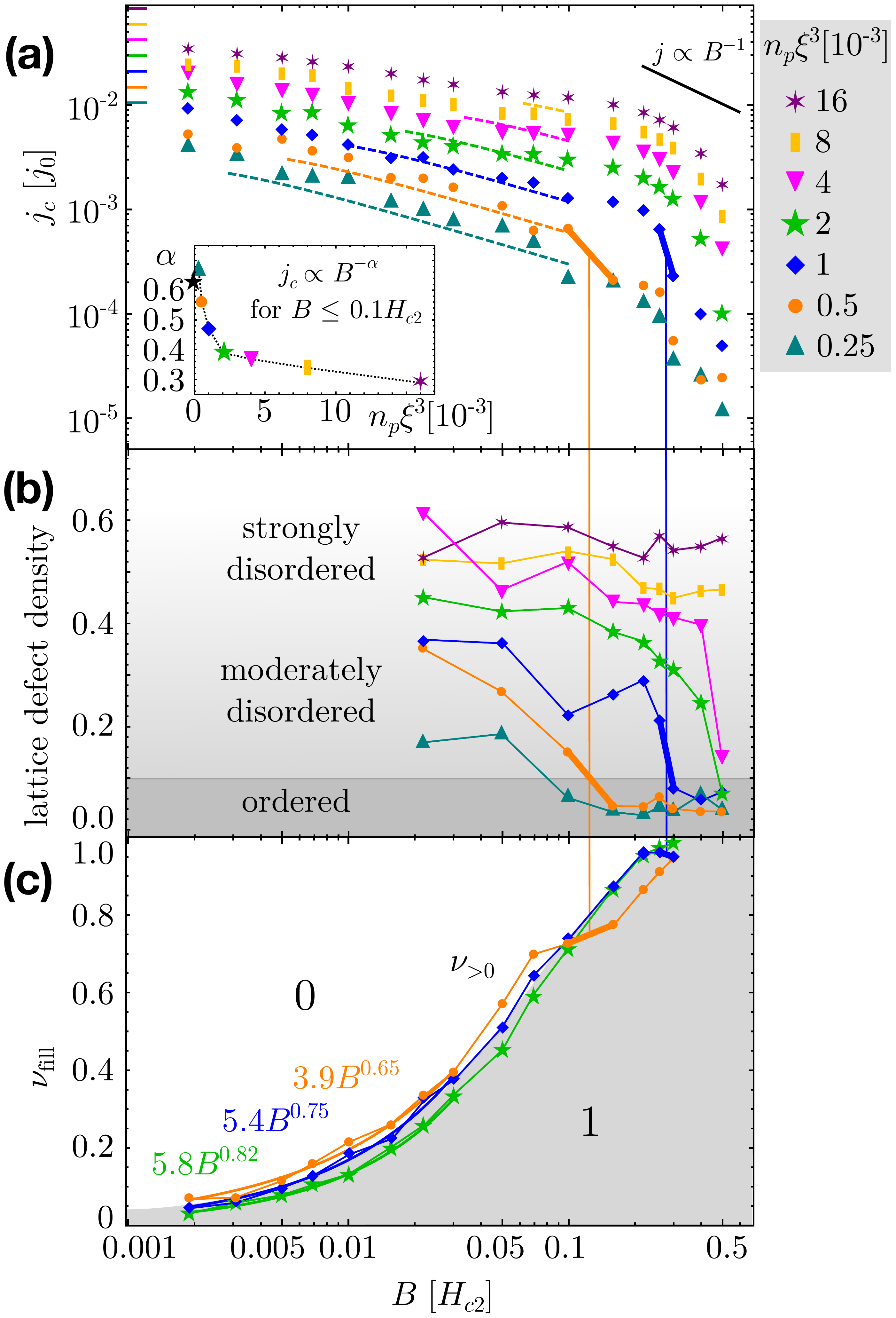}\label{fig:a2_jc_B_a}}
	\subfloat{\label{fig:a2_jc_B_b}}
	\subfloat{\label{fig:a2_jc_B_c}}
	\caption{
		\protect\subref{fig:a2_jc_B_a}~Planar projection of the surface 
		plot in Fig.~\protect\subref*{fig:a2_iv_jc_b}, showing $j_{c}(B)$ 
		for selected densities of small inclusions, $a = 2\xi$. Lines show 
		predictions from strong-pinning theory for (i)~the 1D case (solid) 
		below $0.0015 H_{c2}$, see Eq.~\eqref{eq:jc-1D}, (ii)~the 3D case 
		(dashed) for fields $0.003 H_{c2} < B < 0.1 H_{c2}$ 
		[corrected by doublet contributions, see Eq.~\eqref{eq:doublet-correction}], 
		and (iii)~the high-field case (black) where the scaling $B^{-1}$, 
		see Eq.~\eqref{eq:jc-hf}, clearly fails in describing $j_{c}(B)$. 
		Over a wide field range we observe a power-law 
		$j_{c} \propto B^{-\alpha}$ with decreasing $\alpha$ for increasing 
		inclusion density, see inset. In the latter, the black star indicates 
		the predicted value $5/8$ in the limit $n_{p} \to 0$.
		\protect\subref{fig:a2_jc_B_b}~The density of lattice defects 
		quantifies the transition from an ordered to a disordered vortex state. 
		This transition is accompanied by a kink in $j_{c}$ 
		[panel~\protect\subref{fig:a2_jc_B_a}], as well as a plateau 
		in the fraction of filled inclusions 
		[panel~\protect\subref{fig:a2_jc_B_c}], see vertical lines. 
		Notice that at a fixed field strength, the disorder increases 
		as a function of increasing inclusion density.
		\protect\subref{fig:a2_jc_B_c}~Fraction $\nu_{\mathrm{fill}}$ 
		of inclusions capturing (at least) one vortex. 
		This fraction is only weakly dependent on $n_{p}$. 
		The (almost) full occupation of inclusions, 
		$\nu_{\mathrm{fill}} \gtrsim 0.8$, marks a change 
		in the behavior of $j_{c}$ from 3D strong pinning, 
		to the high-field regime, see panel~\protect\subref{fig:a2_jc_B_a}. 
		No double occupation of inclusions is observed 
		for this particle size, i.e., $\nu_{>1} = 0$.
	}
	\label{fig:a2_jc_B}
\end{figure}

The magnetic-field dependences of the critical current for different densities $n_{p}$ are shown in Fig.~\ref{fig:a2_jc_B}. We can make several qualitative observations. The critical current does not saturate at the lowest fields as naively expected from the theory of 1D strong pinning. Estimates for the limiting value (solid horizontal lines at $B < 0.0015 H_{c2}$) indicate that simulations need to be pushed to even lower fields before reaching that saturation.  The intermediate field range is well captured by the 3D strong-pinning result, again augmented by doublet corrections, see Eqs.~\eqref{eq:doublet-correction} and \eqref{eq:jc-comp-with-theory}. At higher field we find a crossover to a new regime characterized by the faster decay of $j_c(B)$. Further analysis shows that in this regime all inclusions are occupied with vortex lines. We find, however, that the critical current clearly deviates from the expected $B^{-1}$ scaling. We attribute this fast decay to two distinct effects. On the one hand, the pin-breaking force acquires a field dependence, reducing the pinning capability of each inclusion upon increasing the magnetic field, see Fig.~\subref*{fig:pin_breaking_force_a} and discussion in Sec.~\ref{sec:high_fields}. On the other hand, the simulations of a single inclusion suggest, see Sec.~\ref{sec:single_inclusion}, that defects of size $a = 2\xi$ reach the Labusch point $\kappa = 1$ near $B = 0.4 H_{c2}$.\footnote{Note that the scaling $j_c\propto n_p$ at field $B = 0.5 H_{c2}$, see Fig.~\ref{fig:a2_jc_np}, shows that the weak collective model does not describe the behavior of the critical current in this field range yet.} When the defect becomes weak, the isolated-defects theory predicts~\cite{BlatterGK2004, Koopmann2004} a fast drop of the critical current $j_{c} \propto (\kappa - 1)^{2}$. In reality, the critical current, of course, does not vanish at $\kappa = 1$ because of defect doublets and collective-pinning effects.

For a fixed number of particles, we empirically note that the critical current follows a power-law $j_{c} \propto B^{-\alpha}$ over a large field range for $B < 0.1 H_{c2}$. The exponent $\alpha$ increases for decreasing particle number from $\alpha = 0.29$ when $n_{p}\xi^{3} = 16 \times 10^{-3}$ up to $\alpha = 0.66$ for $n_{p}\xi^{3} = 0.25 \times 10^{-3}$. The latter is close to the expected value $\alpha = 5/8\approx 0.625$ from the 3D strong-pinning theory, see inset of Fig.~\ref{fig:a2_jc_B}.

The deviations of the exponent from the theoretical value is most likely related to disorder in the vortex arrays. To characterize degree of this disorder, we performed a Delaunay triangulation for selected $xy$ cross sections of the trapped vortex lattice and evaluated the coordination defect density  $(1/N_{v})\sum_{k=1}^{N_{v}} |c_{k} - c_{0}|$, where with $c_{k}$ the coordination number of the $k$th vortex and $c_{0} = 6$ is the coordination number for an ideal lattice. Figure~\subref*{fig:a2_jc_B_b} shows the field dependences of this parameter for several inclusion densities. We can see, surprisingly, that even for very small densities the lattice is already moderately disordered. It transforms into the practically ideal lattice at distinct magnetic field which rapidly increases with the inclusion density. This transformation is accompanied by pronounced downward jump of the critical current, as emphasized by vertical lines in Fig.~\ref{fig:a2_jc_B}. For large defect densities $n_p\xi^3 > 4 \times 10^{-3}$ the vortex arrays remain strongly disordered in the whole field range.

\begin{figure*}[t]
	\centering
	\subfloat{\includegraphics[width=0.9\textwidth]{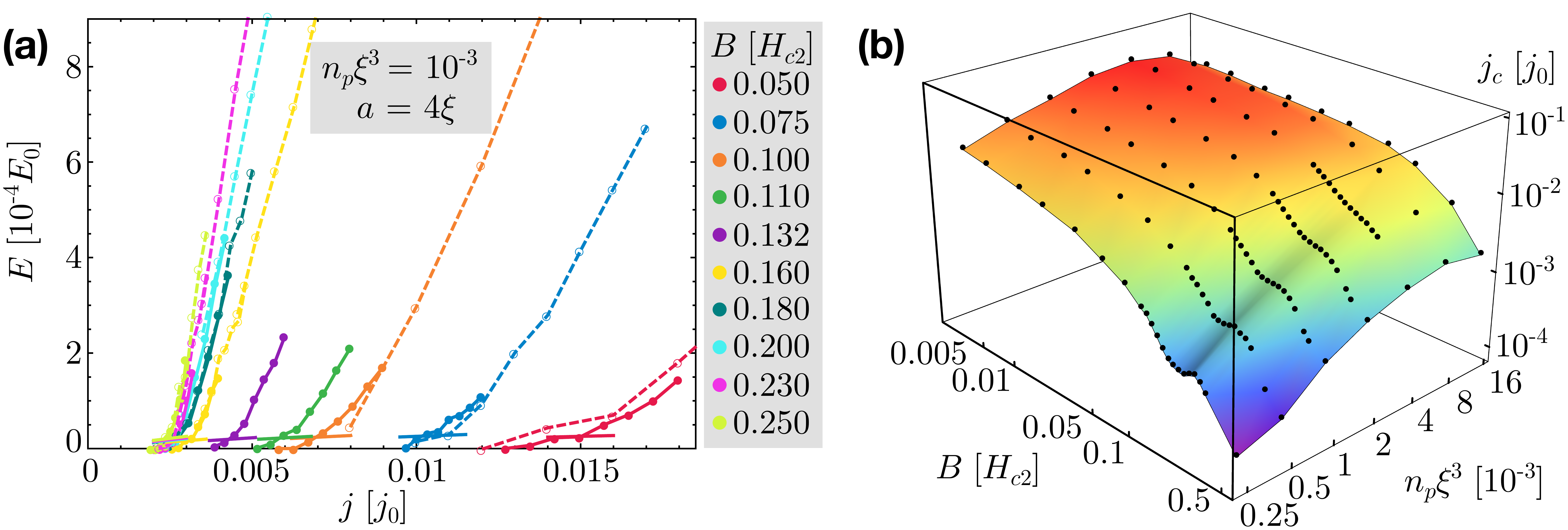}\label{fig:a4_iv_jc_a}}
	\subfloat{\label{fig:a4_iv_jc_b}}
	\caption{
		\protect\subref{fig:a4_iv_jc_a}~Representative set of current-voltage 
		characteristics (at different fields $B$) for a system with fixed inclusion 
		size $a = 4\xi$ and density $n_p = 2 \times 10^{-3}\xi^{-3}$. Open 
		symbols and dashed lines mark \IV dependences obtained using 
		five times faster ramping then ones plotted with closed symbols 
		and solid lines. The short lines indicate the electric fields corresponding 
		to 2\% of the free flux-flow voltage; the intersection with the \IV curve 
		defining the critical current. 
		\protect\subref{fig:a4_iv_jc_b}~Surface plot of the critical current 
		$j_{c}$ for inclusions of size $a = 4\xi$ as a function of their density 
		$n_{p}$ and the applied magnetic field $B$. Black points indicate 
		the $j_{c}$-values used to create the map. For this particle size, the 
		critical current realizes a maximal value at a (field-dependent) optimal 
		density $n_{p}^{\mathrm{opt}}(B) \approx 8 \times 10^{-3} \xi^{-3}$. 
		Note the logarithmic scale on all three axes. The thick frame indicates 
		the direction of the planar projection in Fig.~\ref{fig:a4_jc_B}.
	} 
	\label{fig:a4_iv_jc}
\end{figure*}

Figure~\subref*{fig:a2_jc_B_c} presents the magnetic-field dependence of the occupation fraction $\nu_{\mathrm{fill}}$ of inclusions by vortex lines, for three particle densities, $n_p\xi^3=0.5 \times 10^{-3}$, $10^{-3}$, and $2 \times 10^{-3}$. It should be noted that the occupation fraction weakly depends on the particle density, with only a slight tendency to decrease with increasing $n_p$. Almost all defects become occupied at $B\sim 0.2 H_{c2}$. This field marks the crossover in the $j_c(B)$ dependences in Fig.~\subref*{fig:a2_jc_B_a}. At small fields the occupation fraction grows with field as $B^{\zeta}$, with the exponent $\zeta$ increasing with density from $0.65$ for $n_p\xi^3=0.5 \times 10^{-3}$  to $0.82$ for $n_p\xi^3=2 \times 10^{-3}$, and hence larger than the theoretical value $3/8$ expected in the case of an ordered lattice; remember $\nu_{\mathrm{fill}} \approx S_{t}/a_{0}^{2} \propto a_{0}^{-3/4}$. For $n_p\xi^3 = 0.5 \times 10^{-3}$, a small plateau around $B = 0.1H_{c2}$ is related to the ordering of the vortex lattice in this region. Although weaker, a similar plateau is visible for $n_p\xi^3= 10^{-3}$ near $B = 0.3H_{c2}$. Upon ordering the occupation fraction becomes smaller compared to a disordered configuration.

Having explored the strong-pinning regimes for small particles, we proceed in the next section with a similar analysis for larger inclusions $a = 4\xi$. Studying this defect type---known for its stronger (near optimal) pinning capability---will allow us to embed the current findings in a broader context and to draw comparisons between different defect properties.

\subsection{Large-size particles: role of multiple occupations} \label{sec:a4xi}

In this section we present results for larger spherical particles with diameter $a = 4\xi$ which reveal qualitatively new features, not addressed by a conventional theory. These large inclusions have been explored in a similar field/density range as the small inclusions discussed in Sec.~\ref{sec:a2xi}. Expressed through the volume fraction occupied by the inclusions, $0.008 \leqslant \nu_{\mathrm{vol}} \leqslant 0.39$, the explored range is however significantly different than that for $a = 2\xi$. Snapshots of order-parameter isosurfaces and extracted vortex lines are illustrated in the right column of Fig.~\ref{fig:vortex_configs}. In contrast to the case of smaller inclusions, the vortex arrays remain strongly disordered almost in the whole studied parameter range. Figure~\ref{fig:a4_iv_jc} shows representative series of current-voltage characteristics for a system with $N_p = 500$ inclusions ($n_{p} = 10^{-3}\xi^{-3}$). The dashed lines indicate {\IV} curves obtained from a faster ramping protocol,  with $N_{t} = 5 \times 10^{4}$. Despite the shorter equilibration/average time, the {\IV} dependences are comparable with the ones for slower ramping. Only at low fields, the reduced equilibration time leads to an upwards shift of the current-voltage characteristic. The supplementary data includes several movies illustrating the vortex dynamics for representative magnetic fields at currents slightly exceeding the critical current and concentration $N_p = 500$ ($n_p\xi^3 = 10^{-3}$): for smaller particles with $a = 2\xi$ at \uhref{https://youtu.be/aV4MZzUPYxs}{$B = 0.03 H_{c2}$} and \uhref{https://youtu.be/ae0oME77Pz0}{$B = 0.22 H_{c2}$} well as for larger particles with $a = 4\xi$ at \uhref{https://youtu.be/nTyTnmLOs5o}{$B = 0.03 H_{c2}$} and \uhref{https://youtu.be/YEFLqfXslQs}{$B = 0.22 H_{c2}$}. \uhref{https://youtu.be/P07FsRceVbQ}{Movie} shows the vortex dynamics at applied current significantly larger than the critical current. These and additional movie clips are available at \uhref{https://www.youtube.com/channel/UCjdQ4Ruhxma5pkxGrFxw3CA}{OSCon-SciDAC YouTube channel}.

\begin{figure}[tb]
	\centering
	\includegraphics[width=0.48\textwidth]{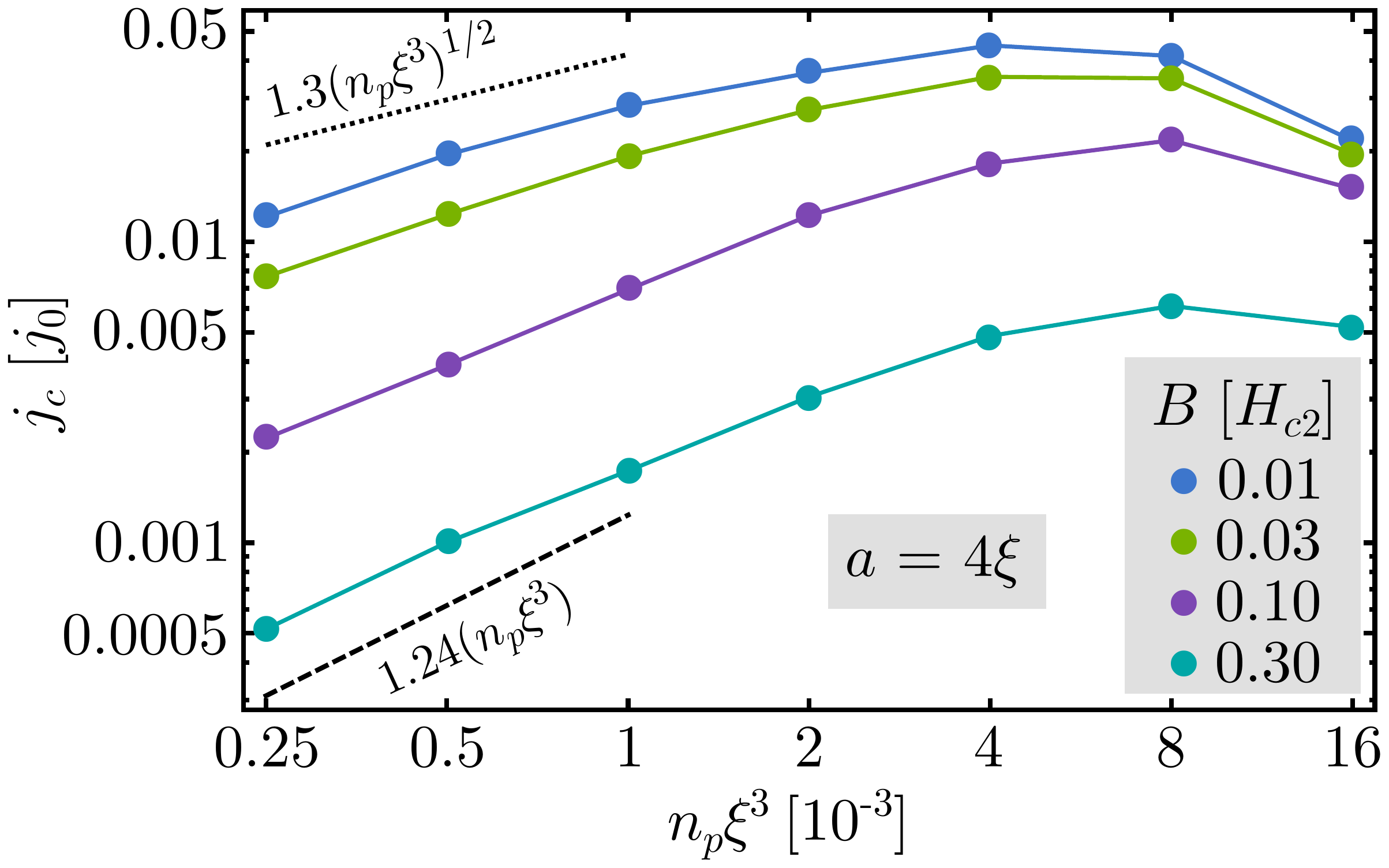}
	\caption{
		The inclusion-density dependences of the critical current 
		for $a=4\xi$ and four magnetic fields.
	}
	\label{fig:a4_jc_np} 
\end{figure}

In Fig.~\ref{fig:a4_jc_np} we present the inclusion-density dependences of the critical current for several magnetic fields. We can see that at small fields $j_c$ grows at small densities approximately as $\sqrt{n_p}$ (1D law), while at high field it grows as $n_p$ (3D law). We also observe that there is a density of inclusions maximizing the critical current $j_c$ at a fixed field. This optimal density slowly increases with increasing $B$, consistent with the results reported in Ref.~\cite{KoshelevPRB16}. Several factors cause a decrease of the critical current at large inclusion's volume fractions~\cite{KoshelevPRB16}. First, vortex lines acquire the ability to jump between neighboring inclusions. Second, a large non-superconducting  volume fraction reduces the effective cross section for the supercurrent leading to an increase of the local current density, a suppression of the order parameter, and, as a consequence, a decrease of the average critical current.

\begin{figure}[tb]
	\centering
	\subfloat{\includegraphics[width=0.50\textwidth]{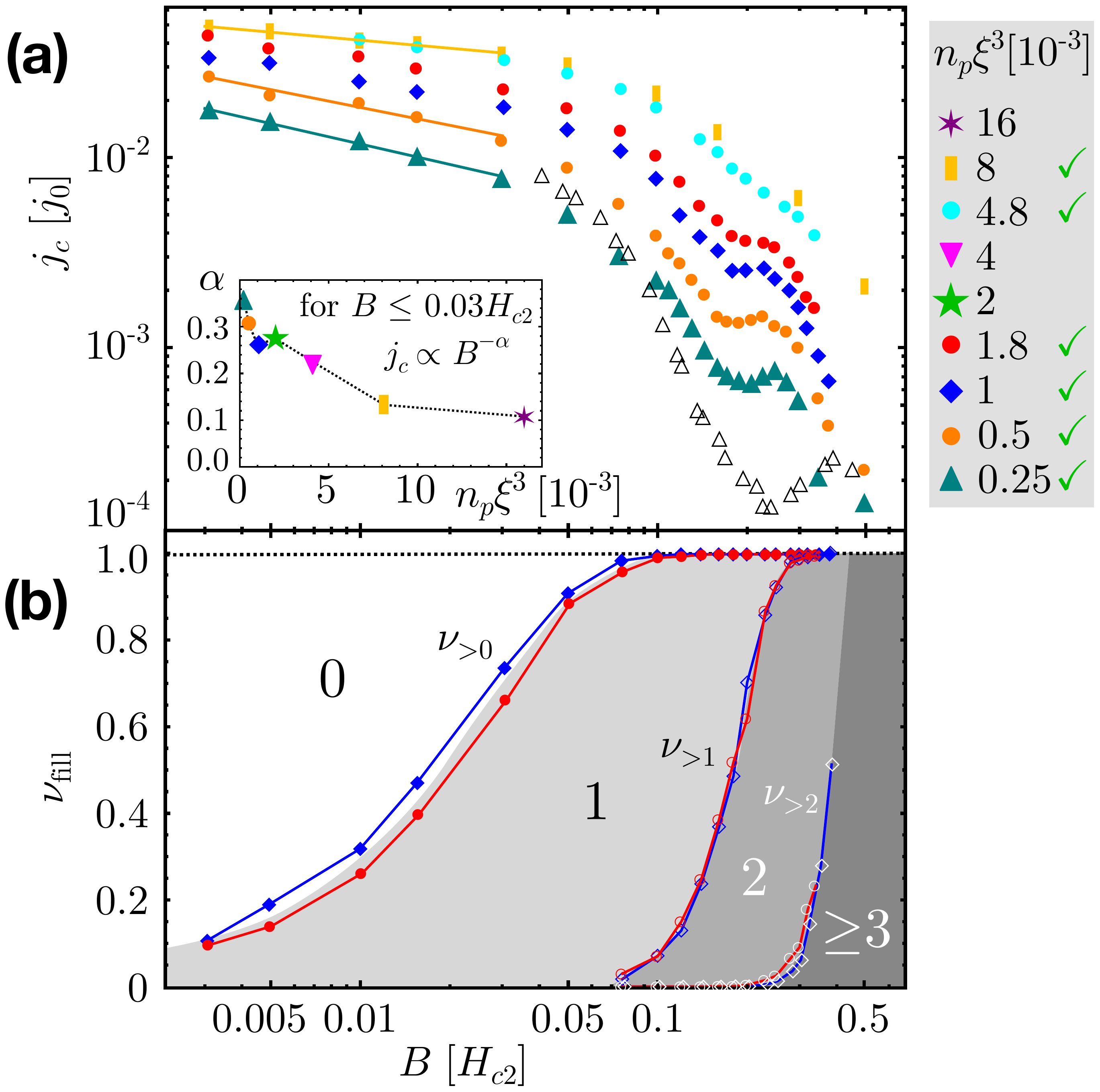}\label{fig:a4_jc_B_a}}
	\subfloat{\label{fig:a4_jc_B_b}}
	\caption{
		\protect\subref{fig:a4_jc_B_a}~Field dependence of the critical 
		current for particles with diameter $a=4\xi$ at different inclusions 
		densities $n_p$ (indicated by check marks). At low fields the critical 
		current obeys a power law behavior $j_{c}(B) \propto B^{-\alpha}$ 
		with a relatively small exponent, see inset. At high fields, when 
		all pin sites are occupied the critical current drops faster. 
		Above a certain field [$\approx 0.2 H_{c2}$, see 
		panel~\protect\subref{fig:a4_jc_B_b}] inclusions can accommodate 
		two vortices. This phenomenon is reflected in a non-monotonic 
		dependence of $j_{c}$ on B, a novel peak effect. Open symbols 
		show a comparison with the single-inclusion results [amplified 
		by the number of inclusions (125) in the system].
		\protect\subref{fig:a4_jc_B_b}~The magnetic-field dependences 
		of pin-occupation fractions ($\nu_{>n}$ means fraction 
		of pins holding more than $n$ vortex lines, 
		$\nu_{>0}\equiv \nu_{\mathrm{fill}}$).
	}
	\label{fig:a4_jc_B} 
\end{figure}

\begin{figure}[tb]
	\centering
	\includegraphics[width=0.42\textwidth]{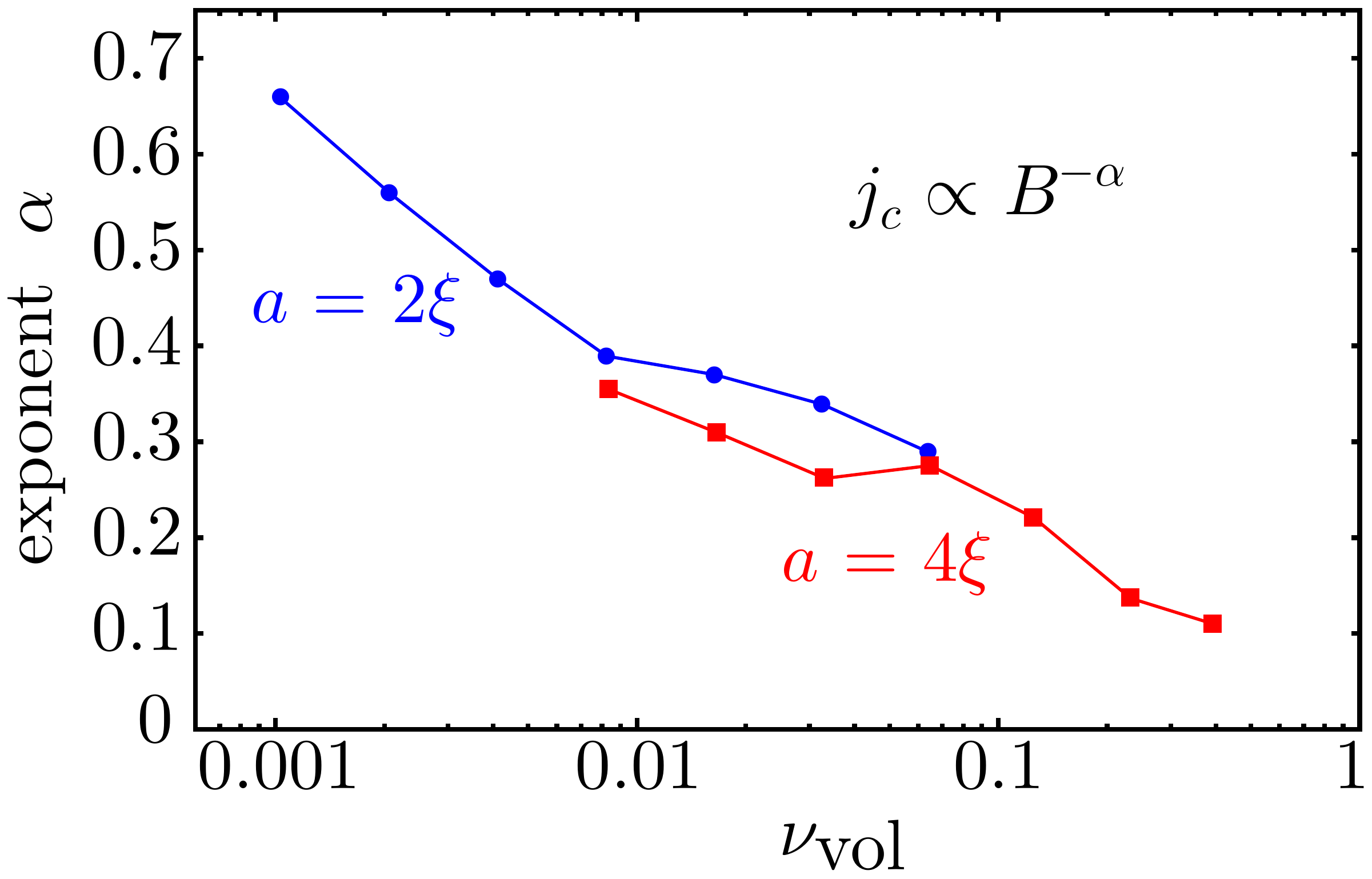}
	\caption{
		Exponent $\alpha$ as a function of the volume fraction occupied 
		by non-superconducting inclusions, 
		$\nu_{\mathrm{vol}} = (\pi/6) n_{p} a^{3} [1 - (\pi/12) n_{p} a^{3}]$. 
	}
	\label{fig:alpha_vf} 
\end{figure}

Figure~\subref*{fig:a4_jc_B_a} shows the magnetic-field dependences of the critical currents for several representative densities. Selected vortex-line configurations for $n_p\xi^3=10^{-3}$ at two fields are shown in Fig.~\ref{fig:vortex_configs}. We identify several distinct regimes. At low fields, $j_c$ decreases slowly. Although not as wide as for $a = 2\xi$, this dependence also may be described by a power-law $j_c\propto B^{-\alpha}$ with $\alpha\approx 0.25$-$0.35$. The exponent slowly decreases with increasing $n_p$. These exponents are similar to ones we found for $a = 2\xi$ in the limit of \textit{large} densities. Even for the smallest particle density the largest exponent $~0.35$ remains significantly smaller than the value $0.625$ suggested by the 3D strong-pinning theory. The plot in Fig.~\ref{fig:alpha_vf} suggests that the exponent $\alpha$ is mostly determined by the \textit{volume fraction} $\nu_{\mathrm{vol}}$ occupied by inclusions. While decreasing at a logarithmic rate $d\alpha/d\log(\nu_{\mathrm{vol}}) \approx -0.2$ for large and small volume fractions, the exponent appears to be weakly dependent on $\nu_{\mathrm{vol}}$ with $\alpha \sim 0.3$ over the wide (and experimentally relevant) range $0.01 < \nu_{\mathrm{vol}} < 0.1$. At intermediate/high fields, the critical current decays faster with an exponent $\alpha > 1$, as in the case of small-size particles. The typical field separating these two regimes slowly grows with the particle density; from $B \sim 0.08 H_{c2}$ for $n_p\xi^3 = 10^{-3}$ to $B \sim 0.12 H_{c2}$ for $n_p\xi^3 = 4.8 \times 10^{-3}$. We attribute this first crossover in the field dependences to full occupation of inclusions with vortex lines, i.e., when $\nu_{1}\equiv \nu_{>0} - \nu_{>1}$ approaches unity, see Fig.~\subref*{fig:a4_jc_B_b}. This figure also illustrates that the occupation of inclusions only weakly depends on $n_{p}$. Faster then $1/B$ decay of $j_c$ in this region is caused by the strong $B$ dependence of the pin-breaking force, see Fig.~\ref{fig:pin_breaking_force}.

At higher fields, $B > 0.15H_{c2}$, we observe a distinct plateau in $j_c$ around $B = 0.2H_{c2}$, evolving into a non-monotonicity at low densities $n_{p}$. One can also see `crowding' of the \IV curves in this field range in Fig.~\subref*{fig:a4_iv_jc_a}. This second crossover and peak effect are clearly caused by \textit{double-occupied} particles. The fraction of such double-occupied inclusions $\nu_{2} = \nu_{>1} - \nu_{>2}$ rapidly grows in the plateau region, changing from $\sim 0.2$ to $\sim 0.9$ in a narrow field range, $0.1H_{c2} < B < 0.23H_{c2}$, see Fig.~\subref*{fig:a4_jc_B_b}. This behavior is in agreement with the single-pin results presented earlier in Sec.~\ref{sec:single_inclusion}. Note that the peak appears far from $H_{c2}$ at a position defined by the defect size; this distinct signature distinguishes the novel peak effect from the classical one. The force with which each inclusion can hold vortices goes down. At the same time, the capability to hold more than one vortex allows to compensate for this effect leading to an upturn in the field-dependence of $j_{c}$. Note that there is no field range of coexistence of unoccupied and double-occupied particles, i.e., the onset of $\nu_2$ at $B / H_{c2} \sim 0.1$ coincides with the saturation of $\nu_{\mathrm{fill}} \to 1$. Similar to the single-occupation fraction $\nu_{1}$, the double-occupation fraction $\nu_{2}$ only weakly depends on the particle density. The $j_{c}$-plateau ends when all particles are at least doubly occupied. Above this field some particles can capture three (or more) vortex lines, and another plateau-like feature may be expected.

In order to draw a direct comparison between the peak effect observed in the critical current of an ensemble of inclusions and the non-monotonic pin-breaking force of a single particle, we have extended the latter to calculate the position-dependent pinning force $f_{\mathrm{pin}}(x)$ and from there its average value $\langle f_{\mathrm{pin}}(x)\rangle$. Substituting $\beta f_{p} = \langle f_{\mathrm{pin}}(x)\rangle$ back into Eq.~\eqref{eq:jc-hf} provides an expression for the critical current at high fields. The result for $n_{p}\xi^{3} = 0.25 \times 10^{-3}$ is shown as open triangles in Fig.~\ref{fig:a4_jc_B}. While the overall trend agrees with the simulation of 125 inclusions, the position of the maximum in the latter case is shifted to lower fields and produces a larger critical current. We attribute both effects to the disordered vortex state for 125 inclusions---as compared to the perfect vortex lattice from the simulations of single inclusions. The disordered state helps pins to capture two vortex lines at lower fields (starting from $0.1 H_{c2}$) and reaches its maximum when all inclusions are doubly occupied (near $0.3 H_{c2}$). Additional quantitative characterizations of the vortex configurations are presented in Appendix~\ref{sec:trapped_vortex_a4}, where we discuss the field dependence of the trapping parameters and mean-square displacements of the vortex lines.

The predicted phenomena of pinning by large inclusions, the strong suppression of the pinning force prior to the onset of defect double occupancy, as well as its revival once the inclusion accommodates two vortices, can be observed in systems with monodisperse particles. A finite distribution of inclusion sizes will lead to a smearing of these effects. For moderately dispersed particles, however, one can expect a plateau in the critical current which indicates the underlying peak effect.

\section{Summary and discussion} \label{sec:discussion}

We systematically investigated pinning properties of randomly distributed spherical inclusions in anisotropic superconductors using large-scale simulations of the TDGL equations. A detailed study and in-depth comparison are presented for two different inclusion diameters, $a=2$ and $4$ coherence length. Our main numerical results can be summarized as follows

\begin{itemize}
\item[--] For both defect sizes we found the intermediate magnetic field regime where the vortex lattice is disordered and a finite fraction of inclusions is occupied with vortex lines. In this regime, the critical current decays with the magnetic field as a power-law $B^{-\alpha}$, where the exponent $\alpha$ decreases with increasing inclusion density (for $a=2\xi$ it drops from 0.66 to 0.3). We found that the exponent $\alpha$ is mostly determined by the volume fraction occupied by the inclusions. 

\item[--] All inclusions become occupied when the magnetic field exceeds a certain value depending on the inclusion size. Above this field, the critical current decreases somewhat faster than the expected $1/B$-law due to the field dependence of the pin-breaking force.  

\item[--]  For $a=2\xi$ and low inclusion densities $n_p$, the lattice becomes ordered at a magnetic field which rapidly increases with $n_p$. The ordering transition---driven by increasing the magnetic field---is accompanied by a reduction of the particle fraction occupied by vortex lines and a sharp drop of the critical current.

\item[--] For large-size particles with $a = 4\xi$, the field dependence is strongly influenced by the occupation of particles with, multiple vortex lines. For small densities, we found a peak in the field dependence of the critical current in the range where the fraction of double-occupied particles rapidly increases with the magnetic field. This peak is smoothed out with increasing particle density. Given that the peak position (as a function of the magnetic field) depends on the defect size only, this feature is clearly distinct from the classical peak effect arising near $H_{c2}$ due to softening of the vortex lattice elasticity.
\end{itemize}

The conventional theory of strong vortex pinning, which we reviewed in Sec.~\ref{sec:estimates}, explains these results only in a very limited range of parameters. We have identified several reasons for this insufficiency. First, in our simulations the vortex lattice is disordered in most of the parameter space. Contrary to our expectations, it only requires a small density of pins to destroy the vortex lattice order. In particular, for the lowest density $n_{p}\xi^{3} = 0.25 \times 10^{-3}$ of small inclusions, the lattice becomes disordered below $B = 0.1 H_{c2}$. This corresponds to $\sim 3.4 \times 10^{-3}$ inclusions per healing volume. Since the strong-pinning theory describes the elastic confinement of a vortex in the lattice by an effective spring constant $\C$, the latter may be significantly altered in the case of a pinned disordered environment. We can expect that the trapping area becomes significantly larger for disordered vortex configurations.

The simplest version of strong-pinning theory suggests that the trapping parameters obey parametric inequalities, i.e., $\up \ll u_{\mathrm{3D}} \ll a_{0}/2$. We find, however, that even for $a = 2\xi$ this situation is only realized at very small magnetic fields. The essential reason for the large trapping lengths is the large anisotropy factor $\gamma = 5$, which enhances both $\up$ and $u_{\mathrm{3D}}$ making them comparable with $a_0$. The generalization of the theory for the case $\up \lesssim u_{\mathrm{3D}} \sim a_{0}/2$ is yet to be done.

Furthermore, our estimates suggest the existence of a wide crossover regime between 1D and 3D limits of strong-pinning theory for which no theoretical description is available. By evaluating the correction to the 3D theory due to close inclusion pairs, we took a first step aiming at closing this gap. With these calculations, we have demonstrated that the existence of such pairs gives rise to \textit{negative} corrections to the critical current $j_{c}-j_{c}^{\sss\mathrm{3D}} = -\eta_{d} n_{p}V_{h}j_{c}^{\sss\mathrm{3D}}$. Including these corrections, the 3D strong-pinning theory reasonably describes our simulation data for $a = 2\xi$ within a finite field range and despite the vortex lattice disorder.

We found that in the crossover regime, the simulated field dependencies of critical currents are well-described by the power-law $\propto B^{-\alpha}$, where the exponent $\alpha$ is smaller that the theoretical value $5/8$ and decreases with increasing inclusion density. Such power-law fall-offs of the critical current are frequently observed in  high-performance superconductors. For REBCO films at low temperatures the exponent $\alpha$ is typically in the range $0.5$--$0.7$~\cite{PolatPhysRevB11, MiuraPhysRevB11, Haberkorn2017}.  Additional defects produced by proton~\cite{JiaAPL13} or oxygen~\cite{LerouxAPL2015} irradiation reduce the exponent (from 0.7 to 0.4--0.5). Such trends are consistent with our simulations, see inset in Fig.~\subref*{fig:a2_jc_B_a}. Another family of materials which typically shows power-law decay of the critical currents is iron-based superconductors. In the pristine crystals of the 122 family, the  exponent is close to $0.5$ indicating strong pinning by some dilute atomic defects~\cite{Fang2012, TaenPRB2012, KihlstromAPL2013, TaenSST2015}. The proton irradiation strongly increases the critical currents and somewhat reduces the exponent (from 0.54 to 0.47 in Ref.~\cite{KihlstromAPL2013}). It was reported that in optimally-irradiated samples, the exponent is close to $0.3$~\cite{TaenPRB2012, TaenSST2015}. Such a decay is consistent with our simulation results in the case of a disordered vortex lattice interacting with a high concentration of particles, see inset in Fig.~\subref*{fig:a2_jc_B_a}. We also found that an exponent close to 0.3 is realized in a wide range of volume fractions occupied by inclusions, namely for $0.01 < \nu_{\mathrm{vol}} < 0.1$ for both inclusion sizes, see Fig.~\ref{fig:alpha_vf}.

The field dependence of the critical currents is noticeable down to the lowest simulated magnetic fields. While this may indicate that the true 1D strong-pinning regime has not yet been reached, we do observe the correct scaling of the critical current with respect to the inclusion density predicted for this regime, $j_{c}\propto \sqrt{n_{p}}$. Such a scaling is expected when vortices wander around to optimize their pinning energy with respect to the elastic line tension while intervortex interactions play a minor role. The field dependence of $j_c$, however, is a clear indicator that these interactions cannot be dismissed in a theoretical description. In our simulations, this field-dependence is further enhanced by the `infinite-$\lambda$' approximation used in the numerical implementation; an approximation which leads to a long-range algebraic decay of intervortex interactions ($\propto 1/r$) at all distances instead of exponential ($\propto e^{-r/\lambda}$). However, a crossover to the 1D regime is expected for the infinite-$\lambda$ model as well.

Finally, the vortex pinning behavior dramatically changes with increasing inclusion size. For large-size inclusions, we have uncovered several new aspects which are not addressed by current theories. First, we have found that at intermediate fields a competition-mediated expulsion of the pinned vortex leads to a very strong field dependence of the pin-breaking force. Second, at higher fields the accommodation of two vortices in the same inclusion leads to a novel peak effect. Note that this peak effect is a property of \textit{monodisperse} defects at small densities. Since the peak position (as a function of field) is determined by the particle size, a realistic situation with a size distribution of inclusions will smooth-out the peak. Nevertheless, given a sufficiently narrow distribution of pinning sites, one may expect a plateau-like feature in the magnetic-field dependence of $j_c$. These new phenomena, occurring at large, near-optimal-size defects urgently call for a generalization of today's theories. 

In conclusion, despite its half-century history, the rich and complex field of vortex pinning still bears many unanswered questions and surprises with new phenomena.

\begin{acknowledgments}
The authors thank L. Civale, V. B. Geshkenbein, W.-K. Kwok, M. Leroux, T. Tamegai, and U. Welp for fruitful discussions. We would like to address a special thank to C. L. Phillips for her technical assistance in extracting the vortex lines from the order-parameter configurations using  the algorithm described in Ref.~\cite{PhillipsPRE2015} and filtering out the field-induced flux lines. The work was supported by the U.S. Department of Energy, Office of Science, Materials Sciences and Engineering Division. A. E. K., I. A. S., and A. G. were supported by the Scientific Discovery through Advanced Computing (SciDAC) program, funded by the U.S. Department of Energy, Office of Science, Advanced Scientific Computing Research and Basic Energy Science. R. W. acknowledges funding support from the Early Postdoc.Mobility fellowship of the Swiss National Science Foundation.
\end{acknowledgments}

\bibliographystyle{apsrev4-1-titles}
\bibliography{pinning_regimes}

\appendix
\onecolumngrid

\section{Corrections to 3D strong-pinning theory due to close pin pairs} \label{sec:doublet}

The theory of 3D strong pinning assumes that each defect acts independently and hence the total pinning force is proportional to their density $n_{p}$. This approximation is justified when the average number of pins within the healing volume $V_{h}$ is small, $n_{p}V_{h} \ll 1$. Even in this case, a random arrangement of inclusions will produce closely located pin pairs (doublets) which do not act independently. This event, occurring with a small probability $\sim(n_{p}V_{h})^{2}$, leads to corrections of the strong-pinning result which we will evaluate in the following.

\begin{figure}[tb]
	\centering
	\includegraphics[width=0.47\textwidth]{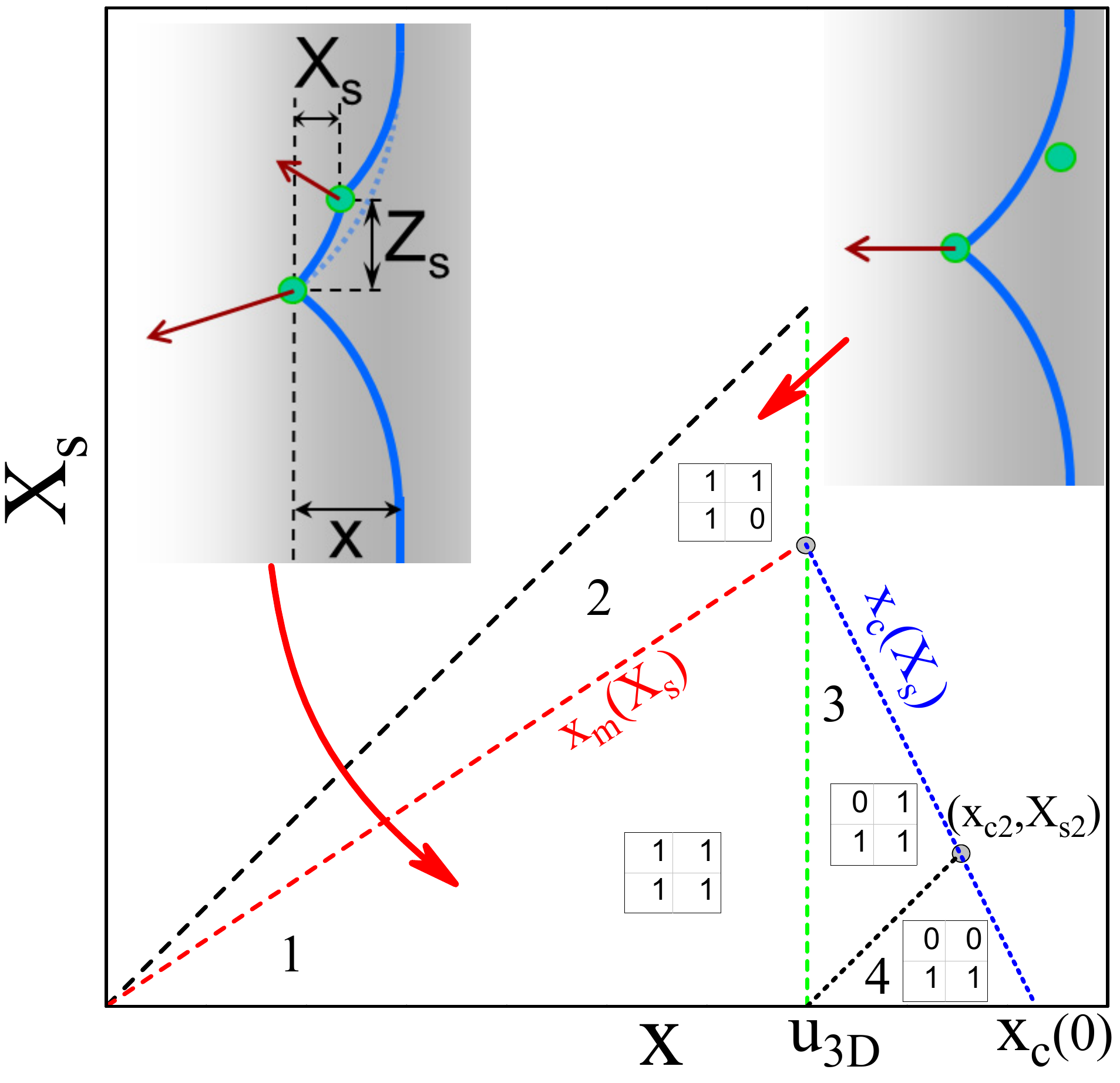} 
	\caption{
		Phase diagram for a pin doublet located in the plane $y = 0$ 
		for a fixed separation $Z_{s}$ along $z$. In each region, 
		the $2 \times 2$ table indicates the occupation of the defects. 
		The first row indicates whether the first (left) or second (right) 
		defect captures the vortex (if acting as isolated pin with 
		coordinates $x$ and $x - X_{s}$); here $1$ stands for occupied 
		and $0$ for unoccupied. The second row shows the same 
		pin occupation for the doublet. Only the regions where 
		$\delta f_{d} \neq 0$ are considered.
	}
	\label{fig:doublet_diagram}
\end{figure}

Starting from Eq.~\eqref{eq:Corrjc}, we obtain the correction to the bulk pinning force
\begin{equation}
	\delta F_{c} = 2\frac{B}{\Phi_{0}}n_{p}^{2} \!\int\! d^{2}\vec{r} \!\int\! d^{3}\vec{R}_{s} \bigl[f_{d}(\vec{r},\vec{R}_{s})
	-f_{\mathrm{pin}}(\vec{r})-f_{\mathrm{pin}}(\vec{r}-\vec{R}_{s}^{\perp}) \bigr].
	\label{eq:CorrFp}
\end{equation}
Within the integration space, the two defects may either be occupied or empty. Furthermore the occupation will depend on whether the isolated ($f_{\mathrm{pin}}$) or the doublet ($f_{d}$) contribution is considered. A phase diagram marking the regions with different occupation numbers is shown in Fig.~\ref{fig:doublet_diagram}. Note that, without loss of generality, we have assumed that the second pin is closer to the vortex position; in the particular case $Y_{s} = 0$, this assumption yields $0 < X_{s}$.

In general, the evaluation of $\delta F_{c}$ is rather complicated. In order to provide a quantitative estimate of the effect, we evaluate the above expression for a simple case. In particular, an exact expression shall be derived when the vortex and the two defects lie in the same $xz$-plane, $\vec{r} = (x,0)$, $\vec{R}_{s}^{\perp}=(X_{s},0)$, while the integration over the transverse coordinates $y$ and $Y_{s}$ will only be accounted for approximately. We consider the simplest case of small-size weakly-strong pins, which can be treated within linear elasticity theory, and approximate interaction of the pinned vortex with its surrounding neighbors by a cage potential. The total energy of the vortex line is then cast by
\begin{equation} \label{En-doublet}
	E_{\mathrm{el}}=\int dz\bigg[\frac{\eone }{2}\Big(\frac{du}{dz}\Big)^{2}+\frac{k}{2}u^{2}\bigg],
\end{equation}
where $k$ measures the strength of the cage potential and $u(z)$ denotes the vortex deformation at the height $z$. When pinned at the first defect, the boundary conditions are $u(0)=-x$ and $u(z\rightarrow\pm\infty)\rightarrow 0$. The equilibrium deformation obeys the minimization condition
\begin{equation} \label{Eq-doublet}
	\frac{d^{2}u}{dz^{2}} = \frac{u}{L_{h}^{2}}
\end{equation}
with the healing length $L_{h}=(\eone /k)^{1/2}$. It is straightforward to evaluate the force $f=\eone \left[u'(0_{+})-u'(0_{-})\right]$
(with $u'\equiv du/dz$) with which the vortex line acts on the defect; a force that has to be smaller than the pin-breaking force~$f_{p}$.

Within this framework, the first isolated pin (if trapping the vortex) deforms the flux line as
\begin{equation} \label{eq:isolated-pin-sol}
	u(z) = -x \exp(-|z|/L_{h}),
\end{equation}
and exerts a force $f_{\mathrm{pin}}(x) = 2 \eone x/L_{h}$. Analogously, the second pin exerts a force $f_{\mathrm{pin}}(x-X_{s})=2\eone (x-X_{s})/L_{h}$. The criterion for the first and second isolated inclusion being occupied reads $x < u_{\sss\mathrm{3D}} = L_{h}f_{p}/(2\eone )$ and $x-X_{s} < u_{\sss\mathrm{3D}}$ respectively.

The pin doublet may realize the simple state where the vortex line is trapped only by one (the first) defect site. In this case the displacement is given by Eq.~\eqref{eq:isolated-pin-sol}, see region 2 in Fig.~\ref{fig:doublet_diagram} and corresponding inset. The second defect is screened by the first one as long as $X_{s} > X_{m}(x,Z_{s}) \equiv x [1-\exp(-Z_{s}/L_{h})]$; translating into a phase boundary $x_{m}(X_{s},Z_{s}) \equiv X_{s} [1-\exp(-Z_{s}/L_{h})]^{-1}$. In all other cases both pins are occupied, adding another boundary condition $u(Z_{s}) = X_{s}$. In this case the solution reads\footnote{In order to keep the notation simple we have assumed $Z_{s} > 0$.}
\begin{equation}
u(z)=\left\{
\begin{array}{ll}
- (x - X_{s}) \exp[-(z - Z_{s})/L_{h}], & \quad\mathrm{for}\: z > Z_{s}, \\
& \\ 
-x \cfrac{\sinh[-(z-Z_{s})  /L_{h}]} {\sinh(Z_{s}/L_{h})} - (x - X_{s}) \cfrac{\sinh(z/L_{h})}{\sinh(Z_{s}/L_{h})}, & \quad\mathrm{for}\: -Z_{s} < z < 0, \\ 
& \\ 
- x\exp(z/L_{h}), & \quad\mathrm{for}\: z < 0.
\end{array}
\right.
\label{eq:Displ2occ}
\end{equation}
Evaluating the condition for depinning from the first pin, we arrive at
\begin{equation}
	\frac{\eone}{L_{h}} \biggl[ x + \frac{x\cosh\left(Z_{s}/L_{h}\right)-(x-X_{s})}{\sinh(Z_{s}/L_{h})} \biggr] = f_{p}
\end{equation}
gives the critical distance, see Fig.~\ref{fig:doublet_diagram}, 
\begin{equation}
	x_{c}(X_{s},Z_{s}) = \bigl[1 + \exp(-Z_{s}/L_{h}) \bigr] u_{\sss\mathrm{3D}} 
	-\cfrac{X_{s}}{\exp\left(Z_{s}/L_{h}\right)-1} 
\end{equation}
or, inversely, $X_{s,c}(x,Z_{s}) \equiv 2\sinh\left(Z_{s}/L_{h}\right)u_{\sss\mathrm{3D}}-\left[\exp\left(Z_{s}/L_{h}\right)-1\right]x$. It is interesting to observe that the presence of the second pin \textit{increases} the critical distance in comparison with an isolated pin, $x_{c} (X_{s}, Z_{s}) > u_{\sss\mathrm{3D}}$, and $x_{c}(X_{s},Z_{s})=u_{\sss\mathrm{3D}}$ at $X_{s}=X_{m,c}(Z_{s})=X_{m}(u_{\sss\mathrm{3D}},Z_{s})$. Correspondingly, in the region $u_{\sss\mathrm{3D}}<x<x_{c}(X_{s},Z_{s})$, shown as regions 3 and 4 in Fig.~\ref{fig:doublet_diagram}, both defects contribute to the pinning force $f_{d}$ of the doublet while at least one isolated pin is unoccupied. In the region 4, i.e., beyond the line $X_{s} = x-u_{\sss\mathrm{3D}}$ both isolated pins are unoccupied. This boundary intersects with the critical line $X_{s,c}(x,Z_{s})$ at $X_{s} = X_{s,2} \equiv [1-\exp(-Z_{s}/L_{h})]\exp(-Z_{s}/L_{h}) u_{\sss\mathrm{3D}}$, defining $x_{c,2} = X_{s,2} + u_{\sss\mathrm{3D}}$.

Evaluating the displacement derivatives $u'(z)$ at both defect heights, one arrives at an expression 
\begin{equation}
	f_{d}(x;X_{s},Z_{s})=\frac{2\eone }{L_{h}}\frac{2x-X_{s}}{1+\exp(-Z_{s}/L_{h})}.
	\label{eq:frc-doubl}
\end{equation}
for the force acting from the doublet on the vortex line. In the most generic region 1, see Fig.~\ref{fig:doublet_diagram}, where all three constituting terms of $\delta f_{d}$ are non-zero, we find
\begin{equation}
	\delta f_{d}(x,X_{s},Z_{s})=-\frac{2\eone }{L_{h}}\frac{2x-X_{s}}{\exp(Z_{s}/L_{h})+1}.
	\label{eq:frcCorr1}
\end{equation}
In order to proceed, we decompose the correction to the bulk pinning force, Eq.~\eqref{eq:CorrFp}, into 
\begin{equation}
	\delta F_{c} = \frac{B}{\Phi_{0}}n_{p}^{2}f_{p}u_{\sss\mathrm{3D}}^{2} \!\int\! dy \!\int\! dY_{s} \!\!\int\limits_{-\infty}^{\infty}\!\!dZ_{s}\mathcal{J}(y,Y_{s},Z_{s}), 
	\label{eq:dFpInt}
\end{equation}
with
\begin{equation}
	\mathcal{J}(y,Y_{s},Z_{s})  = \!\int\! dx \!\int\! dX_{s} \bigl[ f_{d}(\vec{r},\vec{R}_{s})
	- f_{\mathrm{pin}}(\vec{r})-f_{\mathrm{pin}}(\vec{r}-\vec{R}_{s}^{\perp}) \bigr].
\end{equation}
Using the above results, we accurately calculate $\mathcal{J}^{(0)}(Z_{s})=\mathcal{J}(0,0,Z_{s})$. This two-dimensional integration over $x$ and $X_{s}$ naturally splits into four domains\footnote{It should be noted that the lines separating different regions mark a change in the occupation of either one of the isolated defects or of the doublet state, hence producing a discontinuity in $\delta f_{d}$.} shown in Fig.~\ref{fig:doublet_diagram}, $\mathcal{J}^{(0)}(Z_{s})=\sum_{j}\mathcal{J}_{j}^{(0)}(Z_{s})$, where each contribution $\mathcal{J}_{j}^{(0)}(Z_{s})$ is given by
\begin{align}
	\mathcal{J}_{1}^{(0)}(Z_{s}) & = -\cfrac{2}{u_{\sss\mathrm{3D}}^{3}}\!\!\int\limits _{0}^{u_{\sss\mathrm{3D}}}\!\!dx \!\!\!\!\!\!\!\!\! \int\limits_{0}^{\ \ \ X_{m}(x,Z_{s})} \!\!\!\!\! \!\!\!\! dX_{s} \cfrac{2x-X_{s}}{\exp(Z_{s}/L_{h})+1},\\
	\mathcal{J}_{2}^{(0)}(Z_{s}) & = -\cfrac{2}{u_{\sss\mathrm{3D}}^{3}}\!\!\int\limits _{0}^{u_{\sss\mathrm{3D}}}\!\!dx \!\!\!\!\! \int\limits_{X_{m}(x,Z_{s})}^{x} \!\!\!\!\! dX_{s}\ (x-X_{s}),\\
	\mathcal{J}_{3}^{(0)}(Z_{s}) & = \cfrac{2}{u_{\sss\mathrm{3D}}^{3}}\int\limits_{3}dX_{s}\ dx\ \bigg[\cfrac{2x-X_{s}}{1+\exp(-Z_{s}/L_{h})} - (x - X_{s})\bigg],\\
	\mathcal{J}_{4}^{(0)}(Z_{s}) & = \cfrac{2}{u_{\sss\mathrm{3D}}^{3}}\int\limits _{4}dX_{s}\ dx\ \cfrac{2x-X_{s}}{1+\exp(-Z_{s}/L_{h})}.
\end{align}
Here, we have used
\begin{equation*}
	\int_{3}dX_{s}dx \equiv \int_{0}^{X_{s,2}}dX_{s}\int_{u_{\sss\mathrm{3D}}}^{u_{\sss\mathrm{3D}}+X_{s}}dx + \int_{X_{s,2}}^{X_{m,c}}dX_{s}\int_{u_{\sss\mathrm{3D}}}^{u_{\sss\mathrm{3D}}+X_{s}}dx,
\end{equation*}
as well as
\begin{equation*}
	\int_{4}dX_{s}dx \equiv \int_{u_{\sss\mathrm{3D}}}^{x_{c2}}dx\int_{0}^{x-u_{\sss\mathrm{3D}}}dX_{s}+\int_{x_{c2}}^{x_{c0}}dx\int_{0}^{X_{s,c}(x,Z_{s})}dX_{s},
\end{equation*}
and the relation $u_{\sss\mathrm{3D}} = L_{h}f_{p}/2\eone$. Performing the integrations, we obtain 
\begin{align}
	\label{eq:J1}
	\mathcal{J}_{1}^{(0)}(Z_{s}) & =-\frac{\zeta(1-\zeta)}{1+\zeta} \Bigl[1+\frac{1}{3}\zeta \Bigr], \\
	\mathcal{J}_{2}^{(0)}(Z_{s}) & =-\frac{\zeta^{2}}{3}, \\
	\mathcal{J}_{3}^{(0)}(Z_{s}) & =\frac{\zeta(1-\zeta)^{3}}{1+\zeta} \Bigl[ 1+\frac{2}{3}\zeta \Bigr], \\
	\label{eq:J4}
	\mathcal{J}_{4}^{(0)}(Z_{s}) & =\frac{\zeta^{2}(1-\zeta)}{1+\zeta} \Bigl[ 2+\zeta-\frac{\zeta^{2}}{3} \Bigr]
\end{align}
with $\zeta \equiv \exp(-Z_{s}/L_{h}).$ Multiple cancellations in the sum $\mathcal{J}(0,0,Z_{s}) = \sum_{j}\mathcal{J}_{j}^{(0)}(Z_{s})$ lead to the remarkably simple result 
\begin{equation}
	\mathcal{J}(0,0,Z_{s}) = -\frac{\zeta^{4}}{3}.
	\label{eq:J0Result}
\end{equation}
While the contributions \eqref{eq:J1}--\eqref{eq:J4} come with different signs, the negative sign of their sum implies that the doublet correction $\delta f_{p}$ \textit{reduces} the overall pinning force. The correction in Eq.~\eqref{eq:dFpInt} can also be cast into the form 
\begin{equation}
	\delta F_{c}=-2\frac{B}{\Phi_{0}}n_{p}^{2}f_{p}L_{h}u_{\sss\mathrm{3D}}^{2}\int dy\int dY_{s}\ r(y,Y_{s})
\end{equation}
with 
\begin{equation}
	r(y,Y_{s})=-\frac{1}{2L_{h}}\int_{-\infty}^{\infty}dZ_{s}\ \mathcal{J}(y,Y_{s},Z_{s}) > 0.
\end{equation}
Evaluating the last expression using Eq.~\eqref{eq:J0Result}, we obtain $r(0,0)=1/12$. Simplifying transverse integration to $\int dy \int dY_{s} \approx 4\up^{2}$, we arrive at the following estimate
\begin{equation}
	\delta F_{c}\approx-\frac{2}{3}\frac{B}{\Phi_{0}}n_{p}^{2}f_{p}L_{h}u_{\sss\mathrm{3D}}^{2}\up^{2}
	\label{eq:rel-corr-3D-doublet}
\end{equation}
for doublet corrections.

\section{Trapping instability into already occupied pin} \label{sec:double_occupation}

\begin{figure}[tb]
	\centering
	\includegraphics[width=0.35\textwidth]{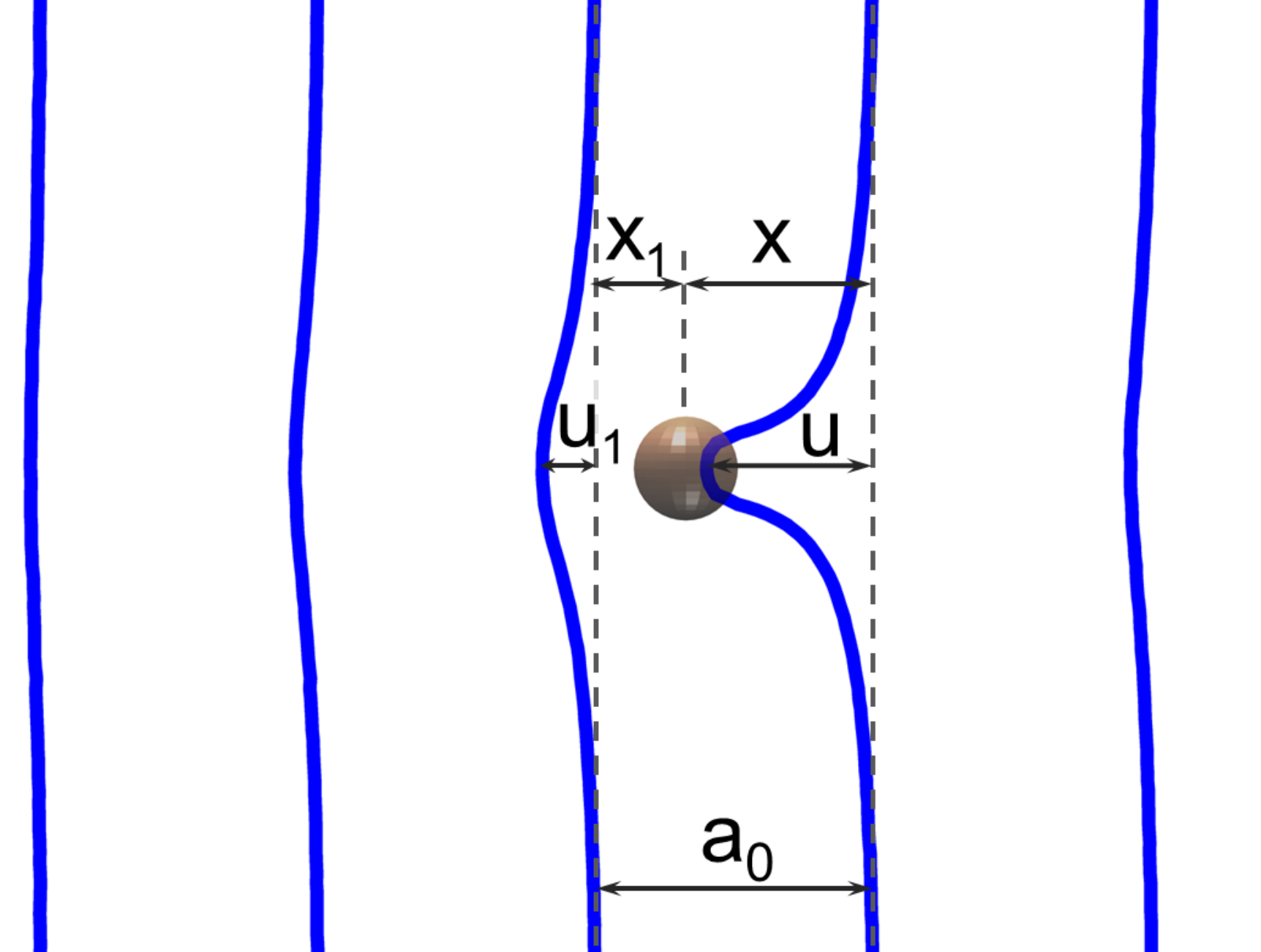}
	\caption{
		The vortex configuration near the double-occupation instability.
	}
	\label{fig:double_occupation_instability}
\end{figure}

Let us consider the situation where two vortices compete for the same defect. Thereby, one vortex shall already occupy the defect while the second vortex is approaching it, see Fig.~\ref{fig:double_occupation_instability}. Asymptotically, the vortices are $a_{0}$ apart. The set of coupled force-balance equations, replacing Eq.~\eqref{eq:non-linearfb}, then read
\begin{align}\label{eq:coupled-1}
	\C u(x) & = f_{p}[x + u(x)] + \Gamma f_{p}[x_{1} + u_{1}(x_{1})],\\
	\label{eq:coupled-2}
	\C u_{1}(x_{1}) & = f_{p}[x_{1} + u_{1}(x_{1})] + \Gamma f_{p}[x + u(x)],
\end{align}
where $x_{1} = x - a_{0} <0$ ($u_{1}$) denotes the asymptotic position (displacement) of the following vortex, and $\Gamma$ measures the reduction in the elastic vortex-vortex interactions at one intervortex distance. Following the route described in Ref.~\cite{BlatterGK2004, Willa2016}, both the effective elasticity $\C = G(0)^{-1}$ and the coupling coefficient $\Gamma = G(a_{0})^{-1}/G(0)^{-1}$ can be expressed through the lattice elastic Green's function $G(r)$. As long as the second vortex is not pinned, i.e. when $|u_{1}| \ll |x_{1}|$, the second term on the right-hand side of Eq.~\eqref{eq:coupled-1} may be neglected. One arrives at
\begin{align}
	\C u(x) & = f_{p}[x + u(x)],\\
	\C [u_{1}(x_{1}) - \Gamma u(x) ] & = f_{p}[x_{1} + u_{1}(x_{1})].
\end{align}
Since $x_{1}$ is coupled to $x$ through the trivial relation $ x_{1} = x - a_{0} < 0$, the second equation may be brought to the form of the first one with $\tilde{u} = u_{1} - \Gamma u$ and $\tilde{x} = x-a_{0} + \Gamma u$. Upon increasing $x$, this second equation reaches an instability at $\tilde{u} = \up$, or
\begin{equation}
	x + \Gamma u(x) = a_{0} - \up
\end{equation}
In the strong-pinning regime, where $u(x) \approx -x$, we find that the instability occurs when $x = (a_{0} - \up) / (1-\Gamma)$. At this point, even if the first vortex has not reached yet its critical deformation $x = u_{\sss \mathrm{3D}} = f_{p}/\C$, the following vortex will get attracted into the defect. As a result, two vortices will co-occupy the same defect. Determining whether the entrance of the second vortex is associated with the immediate departure of the first one or whether the double occupancy of the inclusion is stable requires a separate calculation involving the local repulsion of the two pinned vortices. Phenomenologically, the former scenario will extend over a finite field range after which the defect will be doubly occupied. This picture is validated in the simulations, see Sec.~\ref{sec:single_inclusion}.

\section{Properties of trapped vortex-line configurations for $a=4\xi$} \label{sec:trapped_vortex_a4}

\begin{figure}[tb]
	\centering	
	\subfloat{\includegraphics[width=0.4\textwidth]{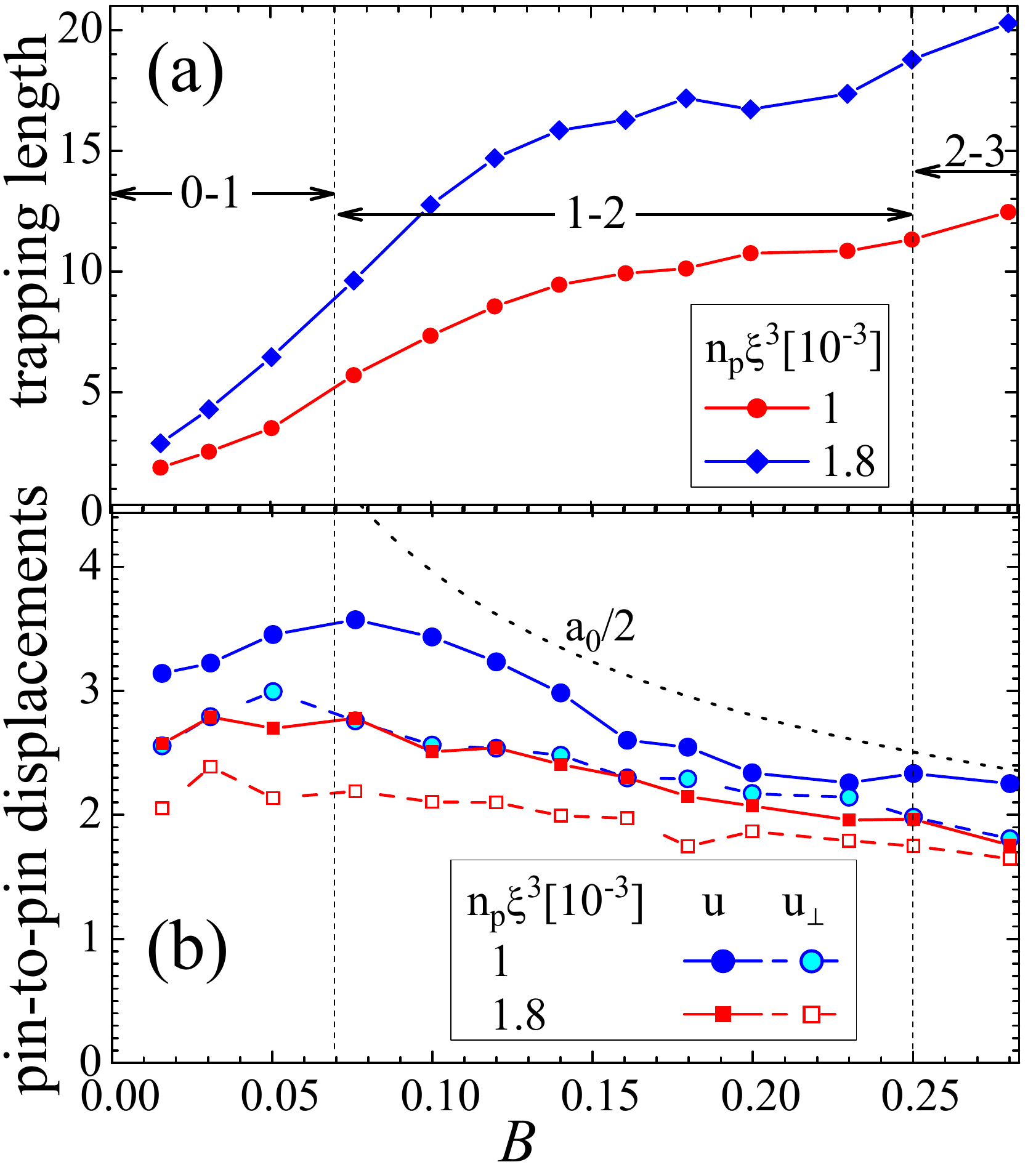}\label{fig:a4_L_B_a}}
	\subfloat{\label{fig:a4_L_B_b}}
	\caption{  		
		Field dependence of parameters characterizing trapped 
		vortex line configurations for two densities of particles, 
		$n_p\xi^3 = 10^{-3}$ and $1.6 \times 10^{-3}$: 
		\protect\subref{fig:a4_L_B_a}~average length of free segments $L$, and 
		\protect\subref{fig:a4_L_B_b}~pin-to-pin displacements $u$ and 
		$\up$, as defined in upper right image of Fig.~\ref{fig:vortices_1d_3d}.
		We marked the regions in which most particles capture either $n$ or 
		$n+1$ vortex lines, with $n=0,1,2$.  
	}
	\label{fig:a4_L_B}
\end{figure}

\begin{figure}[tb]
	\centering	
	\subfloat{\includegraphics[width=0.4\textwidth]{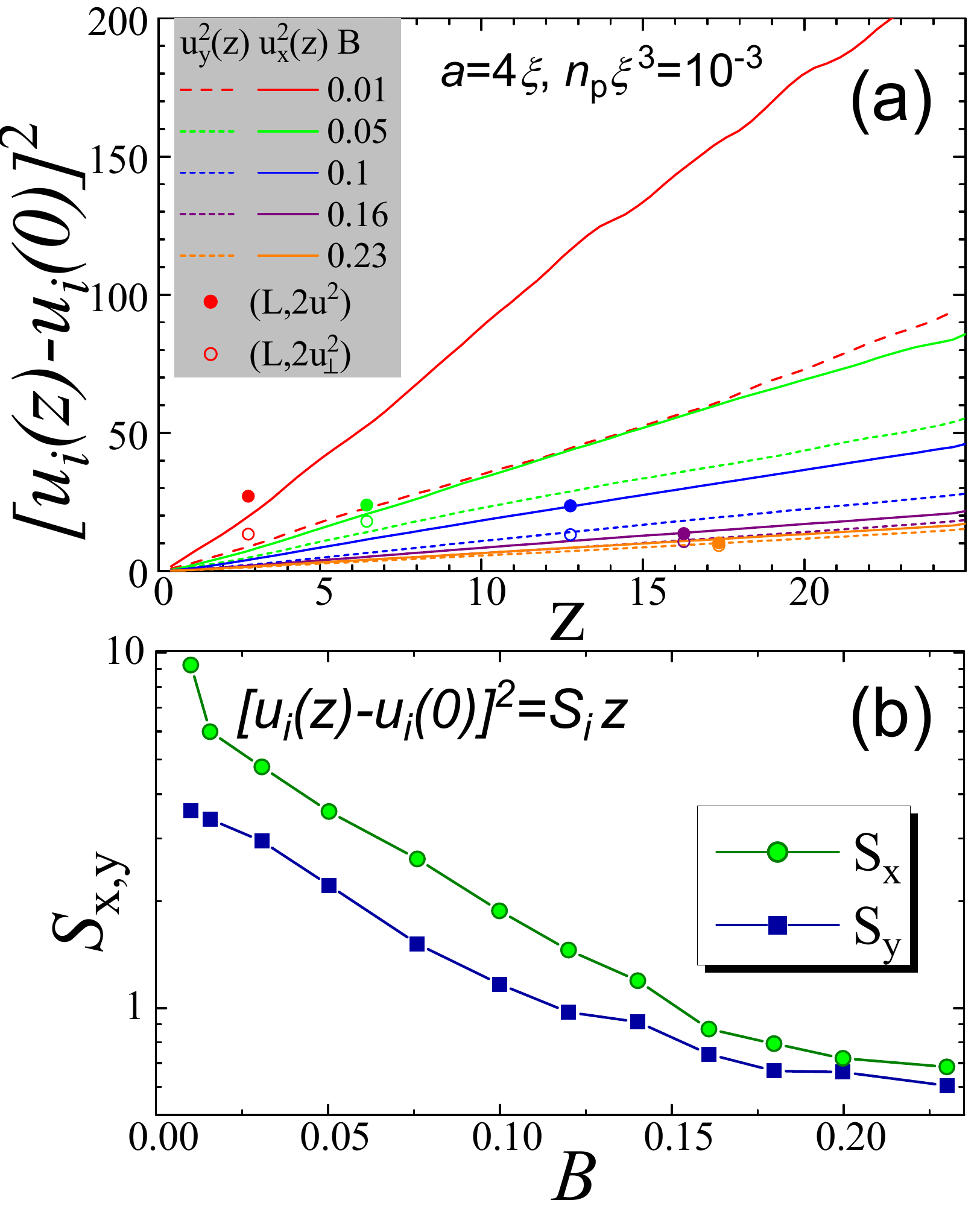}\label{fig:u2_S_B_a}}
	\subfloat{\label{fig:u2_S_B_b}}
	\caption{
		\protect\subref{fig:u2_S_B_a}~Representative mean-squared 
		displacements of the vortex lines for $n_p\xi^3=10^{-3}$ and 
		different magnetic fields. For comparison with the single-segment 
		displacements, we also show points ($L$, $2u^2$) and ($L$, $2\up^2$).
		\protect\subref{fig:u2_S_B_b}~The magnetic-filed dependence 
		of the linear slopes $S_{x,y}$.
	}
	\label{fig:u2_S_B}
\end{figure}

We have performed a parameter characterization of trapped vortex configurations for $a = 4\xi$, see Fig.~\ref{fig:a4_L_B}. The magnetic field dependences are shown for $n_p\xi^{3} = 10^{-3}$ and $1.6 \times 10^{-3}$. Figures \subref*{fig:a4_L_B_a} and \subref*{fig:a4_L_B_b} show the magnetic field evolution of the parameters characterizing geometry of free line segments: their average length $L$ and pin-to-pin displacements $u$ and $\up$, see upper right part of Fig.~\ref{fig:vortices_1d_3d}. The length $L$ grows with the magnetic field in the region of partial occupation of the inclusions up to the crossover field $B / H_{c2} \sim 0.12$. At higher field  the dependence $L(B)$ has a plateau, which is somewhat wider than the similar plateau in the $j_c(B)$ dependence, see Fig.~\subref*{fig:a4_jc_B_a}. The trapping length resume growth when all particles become double-occupied. The pin-to-pin displacements weakly depend on the magnetic field and stay within the range 2--$3.5\xi$, somewhat smaller than the inclusion diameter. As expected, the longitudinal displacement, $u$, is always larger than the transversal one $u_{\perp}$. The difference, however, is not very significant. Surprisingly, the displacements have nonmonotonic field dependence and their maximum is realized roughly at the field of full inclusion occupation. At higher fields, the displacements approximately follow the behavior of the intervortex separation, $a_0$.

In order to characterize the long-range behavior of the vortex lines, we present in Fig.~\subref*{fig:u2_S_B_a} the (longitudinal/transverse) mean-squared line displacement $u_{x,y}^2(z) = \langle[u_{x,y}(z) - u_{x,y}(0)^2]\rangle$ as function of the vertical length $z$, for $n_p\xi^{3} = 10^{-3}$ and different magnetic fields. The displacements in the direction of motion $u_{x}^2(z)$ are always larger than the displacements in the transversal direction $u_{y}^2(z)$. We see that for all magnetic fields the displacements show diffusionlike linear growth $u_{x,y}^2(z)=S_{x,y}z$.  Figure~\subref*{fig:u2_S_B_b} shows the $B$-dependence of the slopes $S_{x,y}$. We can see that the slopes mimic behavior of the trapping length: the rapidly decrease below the crossover field $B / H_{c2}\sim 0.12$ and become field independent at higher fields. Also, at high fields the line wanderings become mostly isotropic $S_x \approx S_y$.

\end{document}